\documentclass[12pt]{article}
\usepackage{amssymb}
\usepackage{latexsym}
\usepackage{amsmath}
\usepackage{color}
\usepackage{cite}
\usepackage{units}
\usepackage{tabularx}
\usepackage{graphicx}
\usepackage{cancel}

\numberwithin{equation}{section}

\newcommand{\lsim}{\raisebox{-0.13cm}{~\shortstack{$<$ \\[-0.07cm] $\sim$}}~}

\begin{document}
\renewcommand{\thefootnote}{\fnsymbol{footnote} }

\pagestyle{empty}
\begin{flushright}
May 2012
\end{flushright}

\begin{center}
{\large\sc {\bf Mitigation of the LHC Inverse Problem}}  

\vspace{1cm}
{\sc Nicki Bornhauser$^{1}$ and Manuel Drees$^{1}$}  

\vspace*{5mm}
{}$^1${\it Physikalisches Institut and Bethe Center for Theoretical
  Physics, Universit\"at Bonn, Nu\ss{}allee 12, D--53115 Bonn, Germany} 
\end{center}

\vspace*{1cm}
\begin{abstract}
  The LHC inverse problem refers to the difficulties in determining the
  parameters of an underlying theory from data (to be) taken by the
  LHC experiments: if they find signals of new physics, and an
  underlying theory is assumed, could its parameters be determined
  uniquely, or do different parameter choices give indistinguishable
  experimental signatures?  This inverse problem was studied before
  for a supersymmetric Standard Model with 15 free parameters. This
  earlier study found 283 indistinguishable pairs of parameter
  choices, called degenerate pairs, even if backgrounds are
  ignored. We can resolve all but 23 of those pairs by constructing a
  true $\chi^2$ distribution using mostly counting observables. The
  elimination of systematic errors would even allow separating the
  residual degeneracies. Taking the Standard Model background into
  account we still can resolve 237 of the 283 ``degenerate''
  pairs. This indicates that (some of) our observables should also be
  useful for the purpose of determining the values of SUSY parameters.
\end{abstract}

\newpage
\setcounter{page}{1}
\pagestyle{plain}

\section{Introduction}
\label{sec:Introduction}

The Large Hadron Collider (LHC) is successfully running and collecting
data. It is hoped that in the near future signs of ``new physics''
will show up. Once a signal for physics beyond the Standard Model (SM)
has been established, one would need to identify the underlying
theory, and to determine its parameters.  For a given theory in an
ideal world a certain parameter choice would lead to a unique
experimental signature. In this case the ``inverse problem'', of going
from experimental observables to parameters of the underlying theory,
would have a unique solution. However, it is quite possible that --
even within a given theory -- several different sets of parameters
reproduce all observables that are available at a given time. Note
that we are here not concerned about (small) regions of parameter
space centered around the true solution; it is clear that in the
presence of non--vanishing experimental errors, a finite region of
parameter space will be allowed even if the inverse problem does have
a unique solution. Rather, the concern is that quite different, not
necessarily connected regions of parameter space cannot be
disentangled using only (future) LHC data.

This issue has been studied in most detail in the framework of the
minimal supersymmetric extension of the SM, the MSSM \cite{mssm}. Most
studies focused on determining the masses of specific sparticles
using features of kinematic distributions, including invariant mass
``edges'' \cite{edges} or ``kinks'' in more complicated observables
\cite{kinks}. This uses only kinematical information, and ideally
allows to directly determine the masses of the involved
superparticles. These can then be compared to predictions of specific
models of supersymmetry breaking, or to fix the parameters of such
models. The model dependence thus enters only at the last stage of the
analysis.

One disadvantage of these kinematical methods is that dynamical
information, i.e. information on counting rates which determine
products of cross sections and branching ratios, is not used at
all. It has been realized quite early that (ratios of) numbers of
events of specific types can be used to discriminate between variants
of the MSSM \cite{early_rate,asym}. More recently, information on the
total event rate after cuts has been shown to improve the performance
of purely kinematical fits \cite{combined}. Even in that case this
method has generally only been used in constrained versions of the
MSSM, with a rather small number of free parameters. This allows to
focus on the most prominent kinematical features, since only a small
number of masses needs to be measured in order to determine all free
parameters of the theory.

Another disadvantage of parameter reconstruction based on kinematic
edges or kinks is that it is not straightforward to automatize
them. Generally human intervention is required to detect an edge. This
method is therefore well suited to detailed analyses of benchmark
points (and eventually of real data, it is hoped), but cannot easily
be used for broad scans of parameter space.

In order to overcome the last two disadvantages, Arkani--Hamed
\textit{et. al.} \cite{bib:inverseProblem} attempted a
``brute--force'' approach to the parameter reconstruction issue in the
context of a quite general version of the MSSM with 15 free parameters
defined at the weak (or superparticle mass) scale.\footnote{A fully
  general MSSM has many more free parameters. However, most of those
  are related to flavor mixing and/or CP violation, and are strongly
  constrained by low--energy observables and/or have little influence
  on collider phenomenology.} They randomly generated 43,026 sets of
parameters, called ``models'' by them. For each parameter set they
Monte Carlo generated the number of events corresponding to an
integrated luminosity of $\unit[10]{fb^{-1}}$ at a center of mass
energy of $\unit[14]{TeV}$. They analyzed these events using a total
of 1808 observables. Most of them are kinematical observables (based
on invariant mass and transverse momentum
distributions\footnote{Rather than looking for edges or kinks in these
  distributions, they bin them in ten or 20 bins, such that each bin
  contains the same number of events. The observables are then the
  boundaries of these bins. These observables can obviously be
  constructed automatically, without human intervention.}); less than
ten percent of the observables are counting observables (number of
events with a certain property). Based on a statistical analysis, 283
so--called degenerate model pairs were identified. These are pairs of
parameter sets that could not be distinguished by their method of
comparison with an estimated $\unit[95]{\%}$ confidence level
\cite{bib:inverseProblem}.

Note that the analysis of ref.\cite{bib:inverseProblem} used a very
large number of observables, but was still unable to distinguish
between even quite different spectra, even though the sparticle mass
scale did not exceed $ \unit[1]{TeV} $. This can be interpreted as implying that
the LHC experiments are in principle incapable of determining all MSSM
parameters in a model--independent fashion.

However, the analysis of ref.\cite{bib:inverseProblem} has several
weaknesses. To begin with, initial state radiation and the
``underlying event'' (thought to arise from interactions between the
``spectator partons'' not taking part in the primary hard interaction)
are ignored. These features enlarge the event; since they do not
depend much on the produced final state, they can be expected to
reduce the observable differences between different parameter sets
even more, i.e. to increase the number of degenerate pairs.

Another criticism is that Arkani--Hamed {\it et al.} use a single
``$\chi^2-$like'' quantity, called $(\Delta S_{AB})^2$, to analyze the
information of all 1808 observables; here $A,\,B$ stand for two sets
of parameters, and $(\Delta S_{AB})^2$ essentially sums the squared
differences between the predicted observables, divided by the squared
total error of these observables, and normalized to the number of
observables included. An observable is included in the definition of
$(\Delta S_{AB})^2$ only if the total error is smaller than both
predictions or smaller than the difference between the predictions;
this is meant to reduce the dilution of $(\Delta S_{AB})^2$ by
observables that are so poorly measured that they cannot discriminate
between parameter sets.

Nevertheless the dilution of the statistical measure by observables
with little discriminatory power remains an issue. In fact,
Arkani--Hamed et al. found \cite{bib:inverseProblem} that 46 out of
the 283 ``degenerate'' pairs had (at least) one observable differing
by more than 5 estimated standard deviations, even if one only
considers the subset of observables constructed from final states
containing two charged leptons. They do not consider this to be
significant, since when comparing 2,600 parameter sets with
themselves, but run with different seed of the Monte Carlo event
generator, they found 611 cases where (at least) one di--lepton
observable differed by more than five estimated standard deviations
between the two runs.

This example casts serious doubt on the estimate of the standard
deviation of $(\Delta S_{AB})^2$ used in
ref.\cite{bib:inverseProblem}. Recall that for a normal (Gaussian)
distributed observable, the probability of two measurements to differ
by more than five (true) standard deviations is about $5.7 \cdot
10^{-7}$. By our count Arkani--Hamed et al. include around 1,000
di--lepton observables. Even if they were all statistically
independent, one would expect at most $5.7 \cdot 10^{-7} \times 1000
\times 2600 \approx 1.5$ observables to differ at more than five
standard deviations. Since the observables used in
ref.\cite{bib:inverseProblem} are actually highly correlated, as
acknowledged by Arkani--Hamed et al., the number of statistically
independent di--lepton observables is much smaller than 1,000, reducing
the probability to observe true $5 \sigma$ fluctuations even more. The
fact that 611 such fluctuations were observed thus indicates a problem
with the estimate of the error and/or with the generation of the
events.

A possibly related issue is that the correlations between observables
are not included in the definition of $(\Delta S_{AB})^2$
\cite{bib:inverseProblem}; this is why this quantity is a
``$\chi^2-$like'' variable, rather than a true $\chi^2$. This means
that the statistical interpretation of this observable is not a priori
clear. Arkani--Hamed et al. address this problem by again comparing
parameter sets to themselves, generated with different seeds of the
random number generator. They found that in $\unit[5]{\%}$ of all
cases $(\Delta S_{AB})^2 > 0.285$ even for the same parameter
set. They therefore defined $(\Delta S_{AB})^2 > 0.285$ as equivalent
to two sets of parameters being distinguishable at $\unit[95]{\%}$
confidence level. This inference is not obvious to us, since the
statistical properties of $(\Delta S_{AB})^2$ are a priori unknown,
and the comparison of parameter sets with themselves yielded results
that seem violently at odds with the usual interpretation in terms of
standard deviations.

Finally, the variable $(\Delta S_{AB})^2$ is constructed to resemble a
$\chi^2$ comparing two different measurements, both of which are
assigned the statistical uncertainties expected for a data sample of
$\unit[10]{fb^{-1}}$. However, the issue is whether, given that Nature chose
parameter set A, a measurement can exclude the {\em prediction} made
for parameter set B. This prediction should (in principle) have
negligible statistical uncertainty. It is not clear to what extent
this distorts the statistics, since the cut--off for $(\Delta
S_{AB})^2$ that defines which pairs are deemed indistinguishable is
determined by Monte Carlo experiments; however, conceptually there is
a significant difference between comparing two experiments and
comparing an experiment with a prediction.

In this paper we re--analyze the degenerate pairs of
ref.\cite{bib:inverseProblem}; for statistical tests we also employ a
larger sample of pairs with slightly larger $(\Delta S_{AB})^2$. We
include initial--state radiation as well as the underlying event, and
analyze events at the hadron level. We want to construct a true
$\chi^2$, in order to have an observable with well--defined
statistical properties. Moreover, we focus on counting signatures. One
reason for this is that it is relatively easy to define statistically
independent counting rates, by simply defining mutually exclusive
classes of events. In contrast, since all events of a given class
contribute to various kinematical observables, these observables will
be statistically correlated. A second reason is that there is a much
larger literature on using kinematical quantities \cite{edges,kinks}
than on the use of counting rates \cite{early_rate,asym,combined}.
Here we want to show that counting rates can play an important role in
discriminating between sets of MSSM parameters; we view this as a
first step towards a determination of the values of these parameters,
including statistically meaningful errors. Altogether we employ (at
most) 84 observables when comparing sets of parameters. Since most of
these observables are counting rates, which should not depend
sensitively on details of the detectors, we do not use a detector
simulation (unlike ref.\cite{bib:inverseProblem}); however, we use
realistic efficiencies when counting $b-$jets and
$\tau-$leptons. Moreover, unlike ref.\cite{bib:inverseProblem} we also
investigate the effect of SM backgrounds.

The remainder of this article is organized as follows. In the next
Section we describe the definition of the sets of parameters, as well
as some technical details of our simulation. In Sec.~3 we list the
observables we use, and construct an overall $\chi^2$ variable out of
them. We perform some checks to show that this variable has the
desired statistical properties. Section~4 contains the results of our
numerical analysis, and we conclude in Sec.~5. Details of the
definition of objects (jets, leptons) and of the cuts employed are
given in the Appendices.

\section{The Simulation}
\label{sec:simulation}

In this Section we give some details of our simulation. We start with
a description of how the sets of MSSM parameters were chosen in
ref.\cite{bib:inverseProblem}, and then describe how we generate events.

\subsection{Parameter Sets}
\label{sec:parameters}

Here we follow ref.\cite{bib:inverseProblem}. There the MSSM spectrum
was parameterized directly at the superparticle mass scale, using the
following 15 free parameters: three gaugino masses $M_1$ (bino), $M_2$
(wino) and $M_3$ (gluino); four independent slepton masses $m_{\tilde
  e_L} = m_{\tilde\mu_L}$, $m_{\tilde e_R} = m_{\tilde\mu_R}$,
$m_{\tilde\tau_L}$ and $m_{\tilde\tau_R}$; six independent squark
masses $m_{\tilde q_{1L}} = m_{\tilde q_{2L}}$, $m_{\tilde q_{3L}}$,
$m_{\tilde u_R} = m_{\tilde c_R}$, $m_{\tilde t_R}$, $m_{\tilde d_R} =
m_{\tilde s_R}$ and $m_{\tilde b_R}$; the Higgs(ino) mass parameter
$\mu$; and the ratio of vacuum expectation values $\tan \beta$. The
masses of the first and second generation sfermions with given quantum
numbers are taken equal; this automatically satisfies stringent
constraints on flavor changing neutral currents in this sector
\cite{mssm}. Four additional parameters are fixed, namely the
trilinear scalar couplings $A_t = A_b = A_\tau = \unit[800]{GeV}$ and
the pseudoscalar Higgs pole mass $m_A = \unit[850]{GeV}$. Out of the
$A-$parameters, only $A_t$ is expected to have significant influence
on LHC phenomenology in the context of the MSSM, but determining its
parameters is expected to be quite difficult. Fixing its value
therefore reduces the difficulty of the overall problem. Similarly,
determining $m_A$ is likely quite difficult, unless it is so small
that the heavy Higgs bosons can be produced abundantly.

43,026 sets of parameters were then generated in
\cite{bib:inverseProblem}, by randomly chosing values of the free
parameters, with probability distributions that are flat within
certain ranges. In particular, the parameters $M_1$, $M_2$ and $\mu$
and the four slepton masses lie between $\unit[100]{GeV}$ and
$\unit[1]{TeV}$. The gluino mass and the six squark masses lie between
$\unit[600]{GeV}$ and $\unit[1]{TeV}$,\footnote{This means that many
  of the scenarios considered in ref.\cite{bib:inverseProblem} are
  likely excluded by published analyses of LHC data
  \cite{lhc_data}. We nevertheless use the same parameter sets in
  order to be able to directly compare our analysis with the results
  of ref.\cite{bib:inverseProblem}.} and $\tan \beta$ varies between
2 and 50. The relations between the parameters are further restricted by the
condition
\begin{equation} \label{bound}
m_{\rm slepton}^{\rm max} < m_{\rm ewino}^{\rm max} + \unit[50]{GeV} <
m_{\rm color}^{\rm max} + \unit[100]{GeV} \, ,
\end{equation}
with $m_{\rm slepton}^{\rm max} $ being the maximum slepton soft mass,
$m_{\rm ewino}^{\rm max}$ being the maximum of $M_1, \, M_2$ and
$\mu$, and $m_{\rm color}^{\rm max}$ being the maximum soft mass or
mass parameter of any color--charged superparticle. The constraints
(\ref{bound}) can be motivated by the fact that most models of
supersymmetry breaking predict \cite{mssm} superparticles with
non--vanishing color charge to be heavier than the color
singlets. They also make it likely \cite{bib:inverseProblem} that
sleptons can be produced in the decay of some color--charged
superparticle, improving the chance that slepton masses can be
determined experimentally.

\subsection{Event Generation}
\label{sec:events}

Given the parameters of the (weak--scale) MSSM Lagrangian, we compute
the supersymmetric spectrum with the program SOFTSUSY
\cite{bib:softsusy}. Next, the branching ratios of kinematically
allowed decays are computed using SUSY--HIT \cite{bib:susyhit}. Signal
events are then generated with the event generator Herwig++
\cite{bib:herwig}. We first generate 10,000 events to determine the
total cross section for the production of superparticles, and then
generate the number of events corresponding to an integrated
luminosity of $\unit[10]{fb^{-1}}$ of data at $\sqrt{s} =
\unit[14]{TeV}$. We include full QCD showering as well as interactions
between spectator partons, using default values of the corresponding
parameters. Since all cross sections are calculated in leading order
(LO) in QCD, we use LO parton distribution functions CTEQ6.6 \cite{pdf}.

We do not attempt any detailed detector modeling. We do not expect
experimental resolutions to be very important for us, since most of
the observables we employ are counting observables, which should be
relatively insensitive to resolution effects. Acceptances (both in
pseudorapidity and transverse momentum) are included in our
definitions of observable leptons and jets, as described in Appendix
A. Moreover, we assume that hadronically decaying $\tau-$leptons as
well as $b-$jets can be tagged with $\unit[50]{\%}$ probability within their
respective acceptance windows. We do not include false positive
tags. Within the MSSM, all flavors of quarks and leptons are produced
with comparable probabilities, so mis--tagging is not likely to
significantly affect the distinction between different parameter sets,
which is at the focus of our analysis.

\section{Method of Comparing Parameter Sets}
\label{sec:method}
\setcounter{footnote}{0}

In this section we describe how we compare parameter sets. To this end
we construct a total $\chi^2$ distribution, which allows us to compute
a $p-$value, which in turn is used to quantify how similar two
parameter sets appear. This is described in the last Subsection. The
first Subsection discusses the observables we use to construct the
overall $\chi^2$, and the second Subsection describes the calculation
of the covariance matrix for these observables.

\subsection{Observables}
\label{sec:observables}

If two different parameter choices yield exactly the same
values for all conceivable observables, then there would be no chance
to distinguish between these parameter sets. In principle this should
not happen for the 15 parameter MSSM we are considering, since each
parameter affects the mass, production cross section and/or decay
branching ratios of at least one superparticle. However, while masses,
cross sections and branching ratios formally are all observables, it
is not clear whether they affect quantities that can actually be
measured by LHC experiments sufficiently strongly to allow
determination of the corresponding parameters from LHC data. Moreover,
it is possible that the values of two or more parameters can be varied
simultaneously such that no LHC observable is changed significantly.

Chosing the right set of observables is a quite non--trivial task. On
general grounds one expects that one needs at least one observable for
each free parameter whose value one wishes to determine.\footnote{This
  is not strictly true. If some observable happens to take its
  absolute minimum or maximum, it alone would suffice to determine all
  free parameters. In practice this is not likely to happen for a
  simple observable (a counting rate or kinematical observable for a
  given final state).}  Supersymmetric extensions of the SM are
notorious for allowing many possible signatures \cite{mssm}; however,
not all signatures are viable for all combinations of parameters. This
argues for using not too few observables, in order to make sure that
one is sensitive to all parameters everywhere in parameter space.

Conversely, an observable should show non--trivial dependence on at
least one of the free parameters in order to be useful. If we combine
all observables into a single quantity, adding observables with little
or no discriminatory power can dilute the effect of those observables
that are sensitive to some parameters, reducing the statistical power
of the test.

Another argument against simply using ``all'' observables, as
(essentially) done by Arkani--Hamed et al. \cite{bib:inverseProblem},
is that this makes it difficult to determine a priori the statistical
correlations between the observables. Since our method of comparing
parameter sets is based on an overall $\chi^2$ variable, we need the
full covariance matrix between all observables, including all
correlations. For example, consider two observables which are
correlated and we do not take that into account. If we now compare two
different parameter sets and the observables used for the comparison
are positively correlated (i.e., one tends to become smaller if the
other one does, and vice versa), the parameter sets can look more
similar than they actually are. On the contrary, if these observables
are anticorrelated, ignoring this correlation would lead us to
over--estimate the difference between the compared parameter sets. In
fact, including an observable which is strongly (anti)correlated with
another observable does not increase the actual amount of information
by much, and hence does not significantly improve our chance to
discriminate between different sets of parameters.\footnote{In the
  extreme case there could be a $\unit[100]{\%}$ correlation between
  two (linear combinations of) observables. This would lead to a
  divergence of the inverse correlation matrix $V^{-1}$, because one
  eigenvalue of $V$ would be zero. This divergence can be removed by
  removing one of the observables from the covariance matrix. An
  actual example of this will be described later in this Section.}
Thus using more observables does not automatically improve the results
of a parameter set comparison.

As mentioned in the Introduction, computing the statistical
correlation between a large number of kinematical observables
constructed from the same set of events is difficult. We therefore use
mostly counting observables, with a single kinematical observable per
class of events.

The production of the heavier superparticles can trigger quite lengthy
decay cascades \cite{mssm}, leading to events with several jets and/or
charged leptons. If we classified all events simply by the number
(e.g. 0, 1, 2 or 3 and more), charge and flavor of charged leptons we
would already have $4^6 = 4096$ classes of events. If we defined
additional sub--classes according to the number of non--$b$--jets
(e.g. 0, 1, 2, 3, or 4 and more) and the number of $b$--jets (e.g. 0,
1 or 2 and more) we would end up with $5 \cdot 3 \cdot 4096 = 61440$
different classes. Even adding a single kinematical observable for
each class would double the total number of observables considered.
This large number of observables does not seem to be practical. For
one thing, our event sample consists of typically 25,000 events after
cuts, so most of the classes would be empty.

Instead we consider 84 independent observables. The first one is the
total number of events after cuts. Since nearly all parameter sets
include a stable neutralino as LSP\footnote{The original parameter
  sets of \cite{bib:inverseProblem} include a few examples where the
  LSP is a slepton, in particular the lighter $\tilde \tau$
  eigenstate; even the set of ``degenerate pairs'' contains such
  examples. Scenarios with stable or long--lived charged slepton as
  LSP would be trivial to distinguish from parameter sets where the
  LSP is a neutralino, using stable charged particle searches. We find
  that all scenarios where the LSP is not a neutralino can be
  distinguished from all other scenarios even when only using our
  standard observables listed below. (In all pairs of parameter sets
  we compare, at least one LSP is the lightest neutralino).}, we
always require a sizable missing $E_T$, which greatly suppresses SM
backgrounds. The details of the cuts depend on the number of charged
leptons (meaning electrons, muons and tagged hadronically decaying
taus), as described in Appendix \ref{sec:eventCuts}. On average around
$\unit[30]{\%}$ of all supersymmetric events pass these cuts and we
only consider those events in the following.

We divide the surviving events into twelve mutually exclusive classes
depending on the number, charge and flavor of the measured ``stable''
charged leptons ($l^\pm = e^\pm$ or $\mu^\pm$; $\tau-$leptons decay
within the detector and are thus non--trivial to identify
experimentally). Note that we only include isolated electrons and
muons with pseudorapidity $|\eta| < 2.5$ and transverse momentum $p_T
> \unit[10]{GeV}$, as described in Appendix A.~1. The twelve classes
of events are:

\begin{itemize}
\item[1.] $0l$: Events with no charged leptons
\item[2.] $1l^-$: Events with exactly one charged lepton, with
  negative charge
\item[3.] $1l^+$: Events with exactly one charged lepton, with
  positive charge
\item[4.] $2l^-$: Events with exactly two charged leptons, with total
  charge $-2$ (in units of the proton charge)
\item[5.] $2l^+$: Events with exactly two charged leptons, with total
  charge $+2$
\item[6.] $l^+_i l^-_i$: Events with exactly two charged leptons, with
  opposite charge but the same flavor; i.e. $e^- e^+$ or $\mu^+ \mu^-$
\item[7.] $l^+_i l^-_{j;\ j \neq i}$: Events with exactly two
  charged leptons with opposite charge and different flavor; i.e. $e^-
  \mu^+$ or $e^+ \mu^-$
\item[8.] $l^-_i l^-_j l^+_j$: Events with exactly three charged
  leptons with total charge $-1$. There is an opposite--charged
  lepton pair with same flavor. For example $e^- \mu^- \mu^+$ or $e^-
  e^- e^+$ 
\item[9.] $l^+_i l^+_j l^-_j$: Events with exactly three charged leptons with
  total charge $+1$. There is an opposite--charged lepton pair with
  same flavor. For example $e^+ \mu^- \mu^+$ or $e^+e^-e^+$
\item[10.] $l^-_i l^-_j l^{\pm}_{k;\ k \neq j, i \ for \ +}$: Events
  with exactly three charged leptons with total negative charge,
  i.e. there are at least two negatively charged leptons. There is
  \underline{no} opposite--charged lepton pair with same flavor. For
  example $e^- e^- \mu^+$ or $e^-e^-e^-$ 
\item[11.] $l^+_i l^+_j l^{\pm}_{k;\ k \neq j, i \ for \ -}$: Events
  with exactly three charged leptons with total positive charge,
  i.e. there are at least two positively charged leptons. There is
  \underline{no} opposite--charged lepton pair with same flavor. For
  example $e^+ e^+ \mu^-$ or $e^+e^+e^+$
\item[12.] $4l$: Events with four or more charged leptons
\end{itemize}

Since we assume first and second generation sleptons to be degenerate,
lepton universality will hold for the first and second generation also
in the MSSM; there is thus little sense in trying to distinguish
between electrons and muons. We do, however, distinguish between
positively and negatively charged leptons. Since the initial state at
the LHC is not CP self--conjugate, there is no reason to assume that
$l^+$ and $l^-$ will be produced with equal rate.\footnote{In the
  context of the MSSM, squark decays into charginos can yield positive
  leptons from $\tilde u$ and $\tilde d^*$ decay, and negative leptons
  from $\tilde d$ and $\tilde u^*$ decays. In the absence of CP
  violation a sizable charge asymmetry \cite{asym} can only result
  from first generation squarks, since squarks of higher generations
  will always be created as squark--antisquark pairs. In case of the
  first generation, essentially only $SU(2)$ doublet squarks can decay
  into charginos. The charge asymmetry is therefore obviously
  sensitive to the difference between the masses of $\tilde u_L$ and
  $\tilde d_L$ -- but these masses are related by $SU(2)$ invariance
  \cite{mssm}, and can be taken as equal as far as LHC experiments are
  concerned. The charge asymmetry will nevertheless be sensitive to
  the ratio of squark and gluino masses. Moreover, the branching
  ratios of $\tilde u_L$ and $\tilde d_L$ will have slightly different
  dependencies on the parameters of the chargino sector.} Moreover,
pairs of charged leptons with opposite charge but the same flavor can
originate from the decay of a single neutralino, $\tilde \chi_i^0
\rightarrow l^+l^- \tilde \chi_j^0$ with $j < i$; all other pairs of
charged leptons must come from the decays of two different
(super)particles. In events containing a pair of oppositely charged
leptons it therefore makes sense to distinguish pairs with equal
flavor from those with different flavor. However, since events with
four or more charged leptons are very rare, we do not attempt to
subdivide them any further.

For each of these twelve classes $c \in \{1, 2, \dots, 12\}$ we save
seven observables $O_{i,c}, \ i \in \{1, 2, \dots, 7\}$:
\begin{itemize}
\item $O_{1,c} = n_c/N$: The number of events $n_c$ contained in the
  given class $c$ divided by the total number of events $N$, i.e. the
  fraction of all events contained in a given class
\item $O_{2,c} = n_{c,\tau^-}/n_c$: The number of events in a given
  class $c$ that contain at least one tagged hadronically decaying
  $\tau^-$ divided by the total number of events in this class
\item $O_{3,c} = n_{c,\tau^+}/n_c$: The number of events in a given
  class $c$ that contain at least one tagged hadronically decaying
  $\tau^+$ divided by the total number of events in this class
\item $O_{4,c} = n_{c,b}/n_c$: The number of events in a given class
  $c$ that contain at least one tagged $b-$jet divided by the total
  number of events in this class
\item $O_{5,c} = \langle j \rangle_c$: Average number of non$-b-$jets
  of all events within a given class $c$
\item $O_{6,c} = \langle j^2 \rangle_c$: Average of the square of the
  number of non$-b-$jets of all events within a given class
  $c$\footnote{If event $i$ in the given class contains $N_j^{(i)}$
    non--$b-$jets, then $\langle j^2 \rangle_c = 1/n_c \sum_{i=1}^{n_c}
    \left( N_j^{(i)} \right)^2$.}
\item $O_{7,c} = \langle H_T \rangle_c$: Average value of $H_T$ of all
events within a given class $c$, where $H_T$ is the scalar sum of the
transverse momenta of all hard objects, including the missing $p_T$
\end{itemize}

Jets are reconstructed using the anti$-k_T$ scheme of FastJet
\cite{bib:fastjet}. They have to satisfy $p_T > \unit[20]{GeV}$ and
$|\eta| < 4.8$; $b-$tagging is possible only for jets with $|\eta| <
2.5$ that contain at least one (decay product of a) $b-$flavored
hadron. $\tau-$leptons can be tagged only if they decay hadronically,
are isolated, and their visible decay products satisfy $p_T >
\unit[20]{GeV}$ and $|\eta| < 2.5$. The tagging efficiency for
taggable $\tau-$leptons and $b-$jets are each $\unit[50]{\%}$,
i.e. each taggable $\tau-$lepton or $b-$jet is randomly taken to be
tagged or not with $\unit[50]{\%}$ probability. See Appendix A for
further details. Since we use only a single kinematical observable for
each class, and the classes are all mutually exclusive, we do not have
to deal with correlations between different kinematical observables.

Note the difference between observables 2, 3 and 4 on the one hand,
and 5 and 6 on the other. The former three observables essentially
count all events that contain at least one object of the given type,
while the latter count the (square of the) number of jets per event.
At least for the parameter sets considered here, the number of events
containing a tagged $\tau$ or $b-$jet is usually small, and the number
of events containing two or more such tagged objects is even smaller;
it is therefore sufficient to simply count the number of events that
contain at least one such object.\footnote{$b-$quarks will nearly
  always occur in quark--antiquark pairs, but the probability that
  both members of such pairs are not only taggable, but also tagged is
  rather low. Also, precisely because $b-$quarks almost always occur
  in pairs distinguishing between events with one or two $b-$tags adds
  additional information only if there is a significant number of
  events with more than one $b \bar b$ pair in the final state; such
  events are very rare in our scenarios.} In contrast, most events do
contain several jets. We therefore use two different observables to
characterize the distribution in the number of jets in the different
event classes.

\subsection{Covariance Matrix}
\label{sec:matrix}
\setcounter{footnote}{0}

In order to perform a statistical analysis using these observables we
need their full covariance matrix, including all correlations. This
subsection contains expressions for all non--vanishing entries of this
matrix. 

The first observable was the total number of events after cuts, $N$;
its variance is simply given by
\begin{equation} \label{sn}
\sigma^2(N) = N.
\end{equation}

The next twelve observables are the fractions of events $n_c/N$ that
belong to each class $c$. These twelve observables are not
independent, since $\sum_{c=1}^{12} n_c = N$, i.e. the fractions add
up to unity. This difficulty could be avoided by simply using the
twelve $n_c$ as variables, which are not correlated, and dropping the
total number of events $N$. The reason why we include $N$ and the
$n_c/N$ is that we will later assign a much smaller systematical error
on the event {\em fractions} $n_c/N$ than on the total number of
events $N$. Note also that the event fractions $n_c/N$ are not
correlated with the total number of events $N$, i.e.
\begin{equation} \label{equ:ccN}
{\rm cov} \left( \frac{n_c}{N}, \, N \right) = 0 \ \ \ (c \in
\{1,2,\dots,12\}) \,.
\end{equation}
On the other hand, the covariance between the fraction of events in two
different classes $c$ and $c'$ is nonzero:
\begin{equation} \label{equ:ccc}
{\rm cov} \left( \frac{n_c}{N}, \, \frac{n_{c'}}{N} \right) = \delta_{cc'}
  \frac{n_c}{N^2} - \frac{n_c \, n_{c'}}{N^3} \ \ \ \ (c, \, c' \in
  \{1, 2, \dots, 12\}) \,. 
\end{equation}
The covariance for identical classes ($c = c'$) equals the variance,
i.e. the usual error on the fraction of events in this class. The
covariance matrix consisting of the twelve entries constructed
according to eq.(\ref{equ:ccc}) is not invertible, because the rows and
columns are linearly dependent. This is a consequence of the
constraint $\sum_c n_c = N$ mentioned above. In fact, it should be
clear that it is impossible to create 13 independent observables
$n_c/N$ and $N$ out of twelve measured event numbers $n_c$. The solution
of this problem is to exclude one $n_c/N$ with $n_c \neq 0$ from our
list of observables, and hence from the covariance matrix. So at the
end we are left with 84 observables carrying different information
instead of 85. 

The observables $O_{2,c} = n_{c,\tau^-}/n_c, \, O_{3,c} =
n_{c,\tau^+}/n_c, \, O_{4,c} = n_{c,b}/n_c$ for a class $c$ have the
variances
\begin{equation} \label{equ:cov234}
\sigma^2(O_{i,c}) = O_{i,c} \cdot (1 - O_{i,c}) / n_c \, \ \ \ i \in \{2,3,4\}.
\end{equation}
Observables referring to different classes $c$ are obviously
uncorrelated.\footnote{Note that we are talking about purely
  statistical correlations here. A fluctuation of events in one class
  can obviously not affect the number of events in a different,
  distinct class, nor can it affect the properties of the events in
  this other class. Different observables may very well be
  ``physically correlated'', in that changing the value of some input
  parameter will lead to simultaneous changes in many observables,
  including observables in distinct classes. This kind of correlation
  need not be included when constructing the covariance matrix. It
  does complicate the estimate of errors on the parameters from an
  overall $\chi^2$ fit, which, however, is beyond the scope of this
  paper.}

Within a given class, observable 4 is obviously uncorrelated with
observables 2 and 3. $n_{\tau^-}$ and $n_{\tau^+}$ have some
correlation if there is a sizable fraction of events containing at
least one $\tau^+ \tau^-$ pair. In the simple situation where $\tau-$leptons can only be produced in such pairs and each $\tau-$lepton is
identified, the number of events containing (at least) one $\tau^-$ lepton would obviously be exactly the same as that containing (at
least) one $\tau^+$. In the more realistic case where a $\tau$ has a
finite (rather small) probability $p_\tau$ to be tagged, the
correlation between $n_{\tau^-}$ and $n_{\tau^+}$ scales like
$p_\tau^2$, while the corresponding diagonal entries of the covariance
matrix scale like $p_\tau$. Note that $p_\tau$ includes the geometric
acceptance and the hadronic branching ratio of the $\tau$ in addition
to the $\unit[50]{\%}$ probability of tagging a taggable $\tau$. Moreover, in
reality most $\tau^\pm$ are produced together with an associated
$\tau-$(anti)neutrino rather than with a second $\tau^\mp$, which
dilutes the correlation between observables 2 and 3 even further. We
therefore ignore the correlation between these observables. We will
check the statistical properties of our $\chi^2$ variable later.

The remaining observables can be written as averages over all events
in a given class, $O_{i,c} = \langle o_i \rangle_c$ with $o_5 = j, \,
o_6 = j^2 , \, o_7 = H_T$. Their variances can be calculated directly
from the definition:
\begin{equation} \label{equ:cov567}
\sigma^2(O_{i,c}) = \frac{1}{n_c - 1} \cdot (\langle o_i^2 \rangle_c
- \langle o_i \rangle_c^2) \, \ \ \ i \in \{5,6,7\}. 
\end{equation}
Of these observables, only $\langle j \rangle_c$ and $\langle j^2 \rangle_c$
are correlated within a given class:
\begin{equation}
{\rm cov} (\langle j \rangle_c, \, \langle j^2 \rangle_c) = \frac{1}{n_c - 1}
\cdot (\langle j^3 \rangle_c - \langle j \rangle_c \, \langle
j^2 \rangle_c)\,. 
\end{equation}
Here $\langle j^3 \rangle_c$ is also determined directly from the
(simulated) events.

\subsection{Test Statistics}
\label{sec:teststat}
\setcounter{footnote}{0}

Our method of comparing parameter sets is based on an overall $\chi^2$
variable, defined as:
\begin{equation} \label{equ:chi2AB}
\chi^2_{AB} = \sum\limits_{m,n} (O_m^A - O_n^B) V^{-1}_{mn} (O_m^A -
O_n^B) \, .
\end{equation}
Here $O_n^A$ is the prediction of parameter set $A$ for the $n-$th
observable and $m, \, n$ run over all relevant observables. In general
the sum in eq.(\ref{equ:chi2AB}) therefore includes summation over
event classes $c$ and over the seven kinds of observables $O_{i,c}$
listed in Subsection~\ref{sec:observables}. We will describe later how
we determine which observables are relevant for a given comparison;
this depends on the parameter sets we are comparing.  Finally,
$V^{-1}$ is the inverse of the covariance matrix $V$ of all relevant
observables, with entries
\begin{equation} \label{equ:Vij}
V_{ij} = {\rm cov}[O^A_i, O^A_j] + {\rm cov}[O^B_i, O^B_j] \,.
\end{equation}
This corresponds to adding the errors for parameter sets $A$ and $B$
in quadrature. The calculation of the covariances for a single
parameter set has been described in the previous Subsection.

The main purpose of the present Subsection is to test the statistical
properties of our $\chi^2$ variable. In particular, we want to check
that it indeed follows a proper $\chi^2$ distribution. To that end we
interpret eq.(\ref{equ:chi2AB}) as if we were comparing two distinct
measurements, each with its own statistical uncertainty. We use
samples of simulated signal events corresponding to $\unit[10]{fb^{-1}}$ of
LHC data at $\sqrt{s} = \unit[14]{TeV} $. We do not include any systematic
uncertainties here, since their statistical interpretation is less
clear. $\chi^2_{AB}$ will follow a $\chi^2$ distribution if the
difference between the measurements is entirely due to statistical
fluctuations. We therefore compare two different simulations of the
{\em same} parameter sets, but with different seeds of the random
number generator.\footnote{Strictly speaking a $\chi^2$ distribution
  is defined via the sum of the quadratic differences between Gaussian
  random variables and their means (i.e., $\sum_n \left( O_n -
    \overline{O_n} \right)^2$), not between the members of pairs of
  Gaussian random variables, divided by the true variance, not by
  its estimator. However, if both members of the pair of Gaussian
  random variables whose squared difference we are computing are
  distributed with the same mean and same variance, a true $\chi^2$
  distribution also results for the sum over the squared differences
  between the two random variables, provided the variance is increased
  by a factor of two. The first condition is satisfied since both
  ``data sets'' are generated for the same parameter set. The second
  condition is approximately satisfied since we are normalizing by the
  total covariance. Of course, replacing the true covariance by its
  estimator obtained from data is standard practice in physics.}

If $\chi^2_{AB}$ follows a proper $\chi^2$ distribution, the
probability $p$ of finding a value bigger than the actual one is given
by
\begin{equation} \label{equ:p}
p = \int\limits_{\chi^2_{AB}}^{\infty} f(z, n_d) dz.
\end{equation}
Here 
\begin{equation} \label{equ:f}
f(z, n_d) = \frac {z^{(n_d-2)/2} \, {\rm e}^{-z/2} }
{ 2^{n_d/2} \, \Gamma(n_d/2)}
\end{equation}
is the $\chi^2$ probability density function and $n_d$ is the number
of degrees of freedom, i.e. the number of observables included in the
sum (\ref{equ:chi2AB}). Note that eqs.(\ref{equ:p}) and (\ref{equ:f})
assume that $\chi^2_{AB}$ has been constructed from $n_d$ Gaussian random
variables. 

As noted above, we verify the statistical properties of $\chi^2_{AB}$
by comparing models to themselves. To that end we simulate each model
twice, using different seeds in Herwig++. For simplicity we simulate
exactly the same number of events in both runs.\footnote{Using a
  different seed in general also leads to a slightly different
  estimate of the total sparticle production cross section. Of course,
  this difference decreases if more events are used to estimate the
  total cross section; recall that we estimate this cross section from
  simulations containing 10,000 events, leading to an uncertainty on
  the cross section from Monte Carlo statistics in the percent
  range. This is already much smaller than the physical uncertainty of
  the predicted cross section, e.g. due to missing higher order
  corrections.} Each comparison yields a $p-$value computed from
eqs.(\ref{equ:p}) and (\ref{equ:f}). Performing many such comparisons
leads to a distribution of $p-$values, which should be flat if
$\chi^2_{AB}$ follows a proper $\chi^2$ distribution. This is a
stringent test of our calculation of the covariance matrix.

In order to achieve higher statistics we use a larger sample of
parameter sets than the 384 sets containing the ``degenerate pairs''
claimed in ref.\cite{bib:inverseProblem}. Recall that Arkani--Hamed
\textit{et. al.} called two parameter sets indistinguishable if their
quantity $(\Delta S_{AB})^2 < 0.285$. All model pairs with higher
values of $(\Delta S_{AB})^2$ are said to be distinguishable, with the
degree of distinguishability increasing with increasing $(\Delta
S_{AB})^2$. For the present test we considered all pairs of parameter
sets with $0.285 \leq (\Delta S)^2 < 0.44$, which are still
relatively difficult to distinguish by the criteria of
ref.\cite{bib:inverseProblem}. The upper bound is chosen so that we
get a few thousand additional parameter sets, compromising between
higher statistics and limited computing power. This sample includes
3305 parameter sets forming in 4654 pairs\footnote{The given $(\Delta
  S)^2$ range actually contains 4658 pairs and 3307 parameter sets,
  but with two sets problems occurred simulating them with SUSY--HIT
  followed by Herwig++.} in the given range of $(\Delta S_{AB})^2$.

We generate two statistically independent data sets for each of these
3305 parameter sets. For each pair of data sets corresponding to a
given parameter set we separately look at the total number of events
and at the seven kinds of observables defined for each event class;
the event classes and observables have been described in
Subsection~\ref{sec:observables}. When analyzing the total number of
events after cuts $N$ obviously only a single term contributes to the
definition of $\chi^2_{AB}$ in eq.(\ref{equ:chi2AB}). In all SUSY
scenarios we consider this event number is large enough to use
Gaussian statistics. (This will be the case for all conceivable SUSY
scenarios, including ``no SUSY'', after inclusion of backgrounds.)

In contrast, the remaining observables $O_{i,c}$ defined in
Subsection~\ref{sec:observables} will be approximately Gaussian
distributed only if class $c$ contains a certain minimal number
of events. Recall that $\chi_{AB}^2$ can be expected to follow
a $\chi^2$ distribution only if the observables used in its
construction are Gaussian variables. This is true to a good
approximation only if sufficiently many events contribute to a given
observable.

For example, as already mentioned at the end of Subsection
\ref{sec:observables} the number of events including tagged
hadronically decaying $\tau-$leptons is in general relatively
small. If there is only one identified $\tau$ event in hundred events
of a given class it does not make sense to compare classes containing
only ten events, because most of these classes would contain zero or
at most one event with an identified $\tau$, and Gaussian statistics
would not be applicable. This can be seen from the fact that the
resulting $p-$value from a single comparison of these classes,
computed according to eqs.(\ref{equ:p}) and (\ref{equ:f}), would take
mostly two discrete values, as can be seen from the following
calculation:
\begin{eqnarray} \label{equ:chi_tau}
\chi^2_{AB} & = & \left( \frac{n_{c,\tau^-}^A}{n_{c,A}} -
  \frac{n_{c,\tau^-}^B}{n_{c,B}} \right)^2 \cdot \left[
  \frac{n_{c,\tau^-}^A}{n_{c,A}^2} \cdot \left( 1 - \frac{n_{c,\tau^-}^A}{n_{c,A}}
  \right) + \frac{n_{c,\tau^-}^B}{n_{c,B}^2} \cdot \left( 1 -
    \frac{n_{c,\tau^-}^B}{n_{c,B}} \right) \right]^{-1} \nonumber \\ 
 & \approx & \frac{(n_{c,\tau^-}^A - n_{c,\tau^-}^B)^2}{n_{c,A}^2} \cdot
 \frac{n_{c,A}^2}{n_{c,\tau^-}^A + n_{c,\tau^-}^B} \nonumber \\ 
 & = & \frac{(n_{c,\tau^-}^A - n_{c,\tau^-}^B)^2}{n_{c,\tau^-}^A + n_{c,\tau^-}^B}
\end{eqnarray}
Here $A$ and $B$ refer to the two data sets, which have been generated
using the same parameter set, and $c$ refers to one of our twelve
event classes. The first line follows from eqs.(\ref{equ:cov234}) and (\ref{equ:chi2AB}), and
in the second line we used the approximations $n_{c,A} \approx
n_{c,B}$ and $n_{c,\tau^-}^A, \, n_{c,\tau^-}^B \ll n_{c,A}$. The
combination $n_{c,\tau^-}^A = 0, \, n_{c,\tau^-}^B = 1$ or vice versa
thus leads to $\chi^2_{AB} \approx 1$, while $n_{c,\tau^-}^A =
n_{c,\tau^-}^B = 0$ obviously gives $\chi^2_{AB} = 0$ for this single
observable. Since for small $n_c$ the observable $n_{c,\tau^-}$ most
likely takes the values 0 or 1, the resulting distribution of $p$
computed from this one observable would certainly not be
flat. Instead, there would be a pronounced peak at $p = 1$, since in
many pairs of data sets there would be event classes containing no
identified $\tau-$lepton. Similar remarks apply to the other
observables, although we expect more events to contribute
non--trivially in these cases.

In the following we therefore require a minimum number of events
$n_{i,{\rm min}}$ for a given observable $O_{i,c}$ to be included in
the computation of $\chi_{AB}^2$. The values of $n_{i,{\rm min}}$
depend on the kind of observable (labeled by $i$), as listed in
Table~\ref{tab:nmin}, but are independent of the event class $c$. We
include $O_{1,c}$, the fraction of all events that belong to class
$c$, in the calculation of $\chi^2_{AB}$ as long as at least one of
the two data sets we are comparing contains $n_{1,{\rm min}} = 10$ or
more events; all other observables are only included if both data sets
contain at least $n_{i,{\rm min}}$ events. The reason for this is that
two data sets are obviously quite different, and hence distinguishable
by LHC experiments, if one of them contains many events of a given
class, while the second has none or very few such events. Note that
the total error will then be dominated by the error on the data set
containing many events, and should be (approximately) Gaussian. In
contrast, the other observables will have an approximately Gaussian
error only if both data sets contain sufficiently many events.

\begin{table*}
\centering
\caption{The minimal numbers $n_{i,{\rm min}}$ are listed for our
  seven kinds of observables. For the fraction of all events that
  belong to a given class, $n_c/N$, at least one of the compared data
  sets $A$ and $B$ has to fulfill this condition. For all other
  observables both data sets have to contain at least $n_{i,{\rm
      min}}$ events in a given class $c$ for $O_{i,c}$ to be included
  in the calculation of $\chi^2_{AB}$.}
\vspace*{3mm}
\begin{tabular}{ll}
\hline\noalign{\smallskip}
Observable					& $ n_{i,{\rm min}} $\\
\noalign{\smallskip}\hline\noalign{\smallskip}
$ n_c/N  $				& 10 \\ 
$ n_{c,\tau^-}/n_c $		& 500 \\ 
$ n_{c,\tau^+}/n_c $		& 500 \\
$ n_{c,b}/n_c $				& 50 \\
$ \langle j \rangle_c $ and $ \langle j^2 \rangle_c $	& 50 \\ 
$ \langle H_T \rangle_c $	& 10 \\
\noalign{\smallskip}\hline
\end{tabular}
\label{tab:nmin}
\end{table*}

The chosen minimal numbers $n_{i,{\rm min}}$ are shown in
Table~\ref{tab:nmin}. They were determined by requiring that the
distribution of $p-$values for a given kind of observable is at least
approximately flat. These $p-$values have been obtained by collapsing
the double sum in eq.(\ref{equ:chi2AB}) over the index $i$ labeling
the type of observable to a single term. This allows us to determine
the $n_{i,{\rm min}}$ one by one. The choices listed in
Table~\ref{tab:nmin} imply that between one and twelve classes of
events are used to determine the $p-$value for a given kind of
observable, depending on the observable and on the input parameters.
As expected from our previous discussion, $n_{2,{\rm min}} = n_{3,{\rm
    min}} = 500$ for the fraction of events in a given class
containing at least one identified $\tau^-$ or $\tau^+$ lepton,
respectively, are the highest. For most of the parameter sets we
investigate, the bulk of events are in the three classes with at most
one lepton; as a result, these three event classes tend to contribute
most to the calculation of $\chi^2_{AB}$ after the conditions
summarized in Table~\ref{tab:nmin} are imposed.

We are now ready to show some results of the self--comparison of the
3305 parameter sets considered here. To this end, we show histograms
of $p-$values computed from eqs.(\ref{equ:p}) and (\ref{equ:f}); each
histogram has up to 3305 entries, since exactly two data sets are generated
for each set of input parameters. As noted earlier, if $\chi^2_{AB}$
defined in eq.(\ref{equ:chi2AB}) follows a $\chi^2$ distribution,
these histograms should be (approximately) flat.

\begin{figure*}[t]
\centering
\resizebox{0.5\textwidth}{!}{
\includegraphics{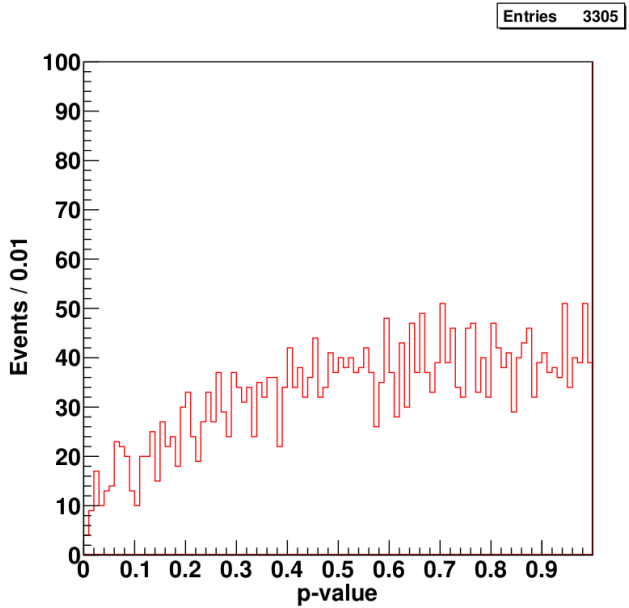}}
\caption{The $p-$value distribution of the self--comparison for the
  sample comprising 3305 parameter sets. Only the total number of events after
  cuts, $N$, is included in the calculation of $\chi^2_{AB}$ and $p$.} 
\label{fig:p_Ntot}
\end{figure*}

Figure~\ref{fig:p_Ntot} shows the distribution of $p-$values if only
the total number of events after cuts $N$ is used in the calculation
of $\chi^2_{AB}$, i.e., $n_d = 1$ in eq.(\ref{equ:f}). Evidently this
distribution is not flat. The bias towards large $p-$values shows that
the compared pairs of data sets are more similar than expected if
$N_A$ and $N_B$ were independent. This is due the fact that for both
seeds of the random number generator the same number of events is
simulated, which is calculated by multiplying the total cross section
for the production of superparticles with the assumed integrated
luminosity of $\unit[10]{fb^{-1}}$. In the absence of cuts we would then
always have $N_A = N_B$, i.e. $p=1$. In practice the number of
accepted events is much smaller than the number of generated events,
and the distribution becomes quite flat for $p \geq 0.4$. Note that we
fix the number of generated events as product of cross section and
luminosity also when comparing different sets of input
parameters. Strictly speaking, we should instead randomly select the
number of generated events from a Poisson distribution whose mean is
given by the product of cross section and luminosity. We do not bother
to do that since this will not change the statistical
distinguishability of two parameter sets significantly. Moreover,
simply fixing the number of generated events is conservative, since it
tends to reduce $\chi^2_{AB}$. Finally, Fig.~\ref{fig:p_Ntot} shows
that the distortion of the $p-$distribution is not very pronounced,
except at very small $p$, once cuts have been imposed.

\begin{figure*}[t]
\centering	
\begin{minipage}[b]{8cm}
\includegraphics[width=8cm]{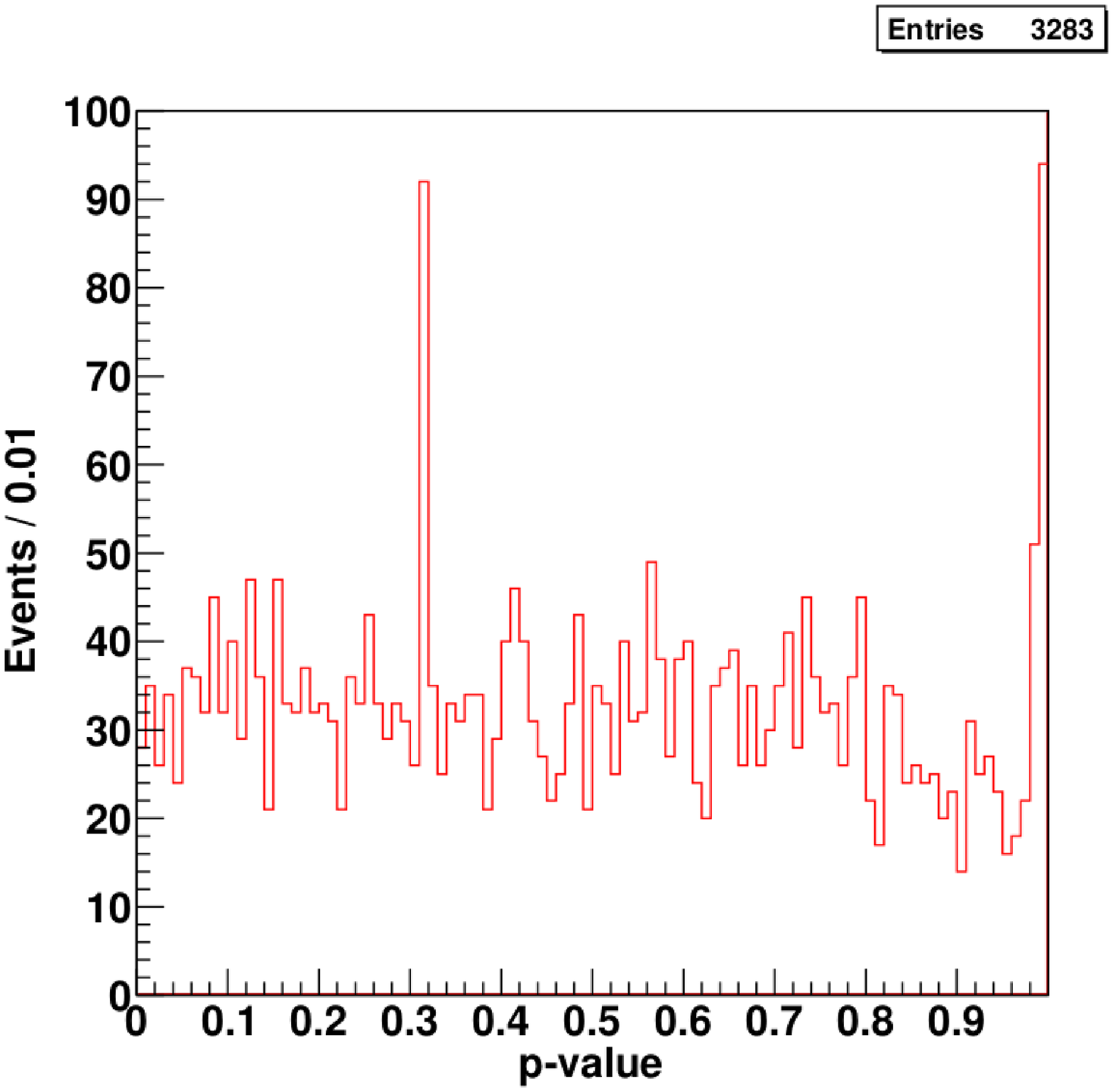}
\end{minipage}
\begin{minipage}[b]{8cm}
\includegraphics[width=8cm]{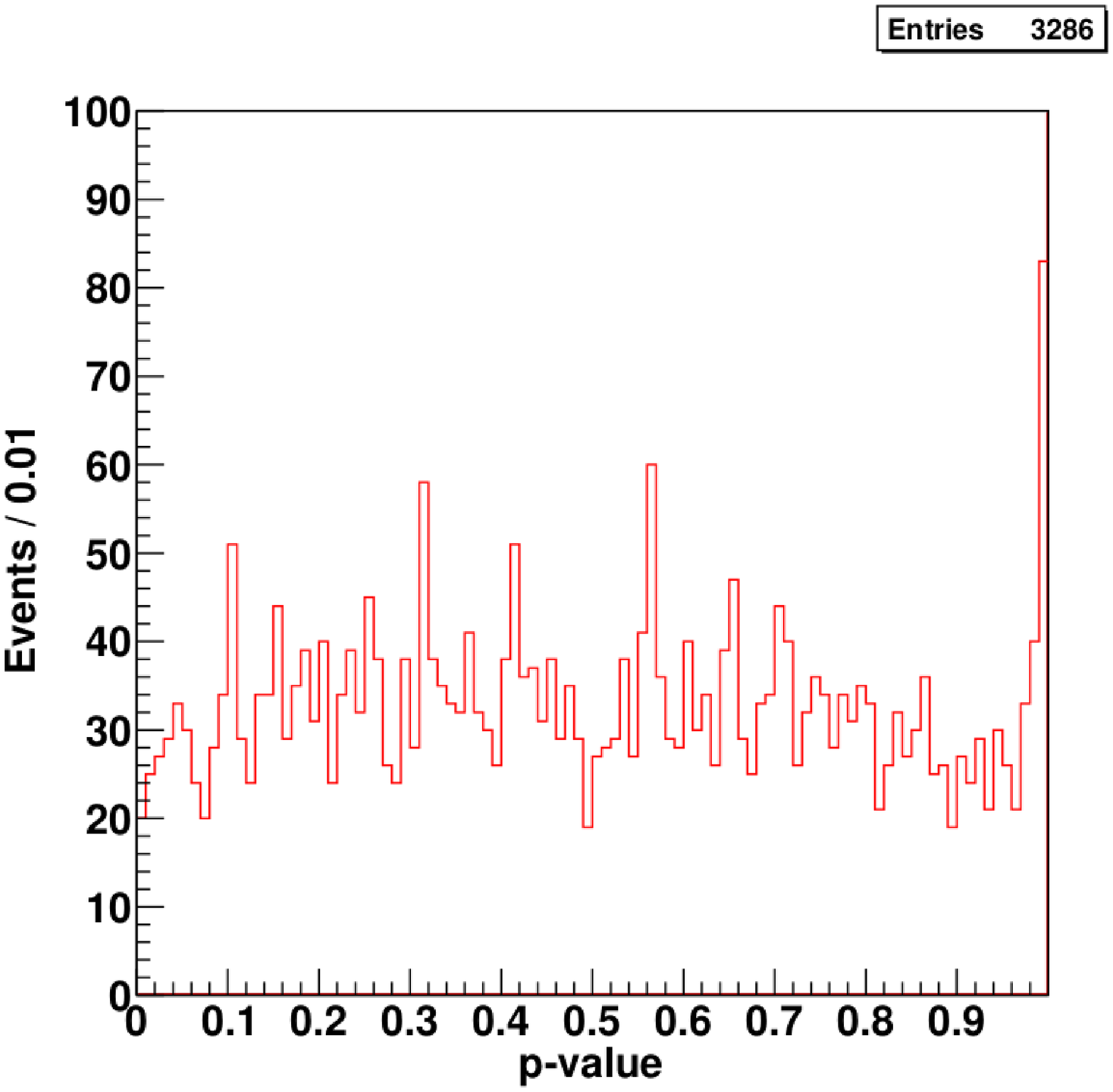}
\end{minipage}
\caption{The $p-$value distribution of the self--comparison for the
  sample containing 3305 parameter sets. Only the fraction of events
  in a given class containing an identified $\tau^-$ (left) or
  $\tau^+$ (right) is included in the calculation of $p$. On average
  2.8 classes satisfy the requirement $n_c \geq 500$ (see
  Table~\ref{tab:nmin}) and are thus included in the definition of
  $\chi^2_{AB}$. The number of entries are slightly smaller than 3305
  since a pair of data sets for a given set of input parameters is
  excluded if all variances (of the considered observables) in the
  compared classes vanish for both data sets or if no class fulfills the $n_c \geq 500$ requirement.} 
\label{fig:p_Ntau}
\end{figure*}

The $p-$distributions for the fraction of events containing a
reconstructed $\tau$ are shown in Fig.~\ref{fig:p_Ntau}. They follow
an overall flat distribution, except for peaks at certain
positions. As discussed above, these peaks are caused by the in
general low number of events containing identified $\tau-$leptons. Since the LHC is a proton--proton collider and therefore has
a net positive charge in the final state, fewer hadronically decaying
$\tau^-$ than $\tau^+$ are identified. Since we use the same minimal
number $n_{2,{\rm min}} = n_{3,{\rm min}} = 500$ of required events in
a given class for both charges, the peaks are more pronounced for
$\tau^-$ events.

For example, if only a single class contributes to the calculation of
$\chi^2_{AB}$, and we have exactly one event with an identified
$\tau^-$ in one data set and none in the other, eq.(\ref{equ:chi_tau})
gives $\chi^2_{AB} \simeq 1$; the corresponding $p-$value for $n_d =
1$ compared observable is $p \approx 0.32$ which is the position of
the first peak. Similarly, the peak at $p = 1$ results from pairs of
data sets where each class contains the same number of identified
$\tau-$leptons of a given charge. This yields $\chi^2_{AB} \ll
1$\footnote{Eq.(\ref{equ:chi_tau}) gives $\chi^2_{AB} = 0$ in this
  case; the exact calculation gives a non--vanishing, but small,
  value, unless the two values of $n_c$ also happen to coincide
  exactly, or $n^A_{\tau} = n^B_{\tau} = 0$.}, and hence $p \simeq
1$. Note that pairs of data sets containing not a single identified
$\tau-$lepton of a given charge in {\em any} event class are excluded
from this comparison since in this case both the numerator and the
denominator in our definition (\ref{equ:chi2AB}) vanish. Additionally some pairs do not fulfill the $n_c \geq 500$ requirement for any class. As a result,
the numbers of entries in the two frames of Figs.~\ref{fig:p_Ntau} are
slightly smaller than 3305. Of course, in this case these observables
cannot be included in the calculation of the overall $\chi^2_{AB}$,
either.

We consider the behavior of the histograms in Fig.~\ref{fig:p_Ntau}
acceptable. Having too many pairs of data sets giving a large
$p-$value, in particular $p=1$, is again conservative. The peak at
$p=0.32$ probably underestimates the true $p-$value for cases where
one data set has one identified $\tau-$lepton with a given charge and
the other has zero\footnote{The best estimate of the true expectation
  value of the number of events with identified $\tau-$lepton is then
  $1/2$, again ignoring the difference in the numbers of events of the
  given class between the two data sets. Using Poisson statistics we
  find that the probability that our $\chi^2_{AB} \geq 1$ is then
  about 0.49.}, but not by very much. Note also that these unwarranted
peaks are due to data sets containing very few identified $\tau-$leptons.
In these cases the $\tau$ observables do not contribute very much to
our total $\chi^2_{AB}$, and hence do not affect the calculation of
the overall $p-$value very much.

\begin{figure*}[t]
\centering
\resizebox{0.5\textwidth}{!}{
\includegraphics{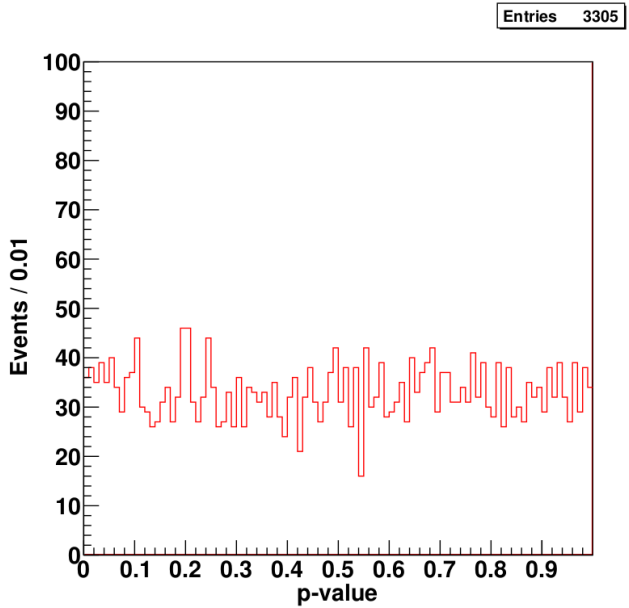}}
\caption{The $p-$value distribution of the self--comparison for the
  sample containing 3305 parameter sets, including all relevant
  observables. On average around 40 observables are compared for a
  pair of data sets.}
\label{fig:p_tot}
\end{figure*}

We checked that the remaining five kinds of observables have flat
distributions (not shown) once the conditions summarized in
Table~\ref{tab:nmin} have been applied. This justifies our choice of
$n_{i,{\rm min}}$ values that define when a given observable is
considered relevant.

Finally, in Figure \ref{fig:p_tot} the overall $p-$distribution is
shown using all types of observables. On average around 40 out of 84
observables are compared for each pair of data sets. The flat shape
confirms our calculation method of $\chi^2_{AB}$, including the
calculation of the covariance matrix as described in
Subsection~\ref{sec:matrix}. It also confirms our expectation that the
anomalies we found above in the $p-$value distributions for the total
number of events after cuts and for the $\tau$ observables do not
distort the overall $p-$value significantly.

We also performed a second kind of statistical test, using the 384
parameter sets that contain the indistinguishable pairs found in
ref.\cite{bib:inverseProblem}. We again performed two simulations for each
parameter set, one corresponding to the nominal luminosity of $\unit[10]{fb^{-1}}$, and one with ten times more events. The first simulation
defines the ``measurements'', while the second simulation was used to
determine the true expectation values of our observables, since their
(statistical) errors are much smaller than those of the
``measurements''. We then checked whether the normalized differences
between the ``measurements'' and their expectation values follow a
Student's $t-$distribution, as they should if the ``measured''
observables obey Gaussian statistics. These tests were also
successful. However, they are somewhat less powerful, since our
computer resources did not permit to generate the required large
number of events for the larger number of parameter sets described
earlier in this Subsection.

\section{Results}
\label{sec:Results}
\setcounter{footnote}{0}

Having verified that our variable $\chi^2_{AB}$ behaves like a proper
$\chi^2$ distribution under appropriate conditions, we are now ready
to use it to analyze whether pairs of parameter sets can be
distinguished by LHC experiments. Our focus will be on pairs of
parameter sets that Arkani--Hamed {\it et al.}
\cite{bib:inverseProblem} deemed to be indistinguishable. Recall that
they found 283 such pairs formed from 384 different sets of
parameters.

As mentioned in the Introduction, two parameter sets $A$ and $B$ can
be considered distinguishable if a measurement can exclude the
predictions made for parameter set $B$, under the assumption that
Nature is described by parameter set $A$, or vice versa. Here the
predictions yield the expectation values of our 84 observables, and do
not have a true statistical uncertainty, given sufficiently large
Monte Carlo statistics for the computation of these expectation
values. Given our limited computer resources, we computed the
predictions using ten times more events than generated for the
measurement; this ensures that the Monte Carlo statistical uncertainty
on the predictions is already almost negligible compared to the
expected statistical uncertainty of the measurement, but we do take
this uncertainty on the prediction into account.

In most cases it should not make a difference if the experiment is
based on parameter set $A$ while the prediction is for parameter set
$B$ or vice versa. This should only be important if the statistical
errors (for fixed integrated luminosity) of $A$ and $B$ differ
significantly. This can only happen if the predictions for $A$ and $B$
are quite different, in which case these parameter sets should be
distinguishable anyway. In order to have a unique comparison rule we
simply symmetrize the expression for the covariance matrix under the
exchange of $A$ and $B$, i.e. we take the average of the covariance
matrix that results when measurements based on $A$ are compared to
predictions based on $B$ and the covariance matrix that describes the
difference between measurements based on $B$ and predictions based on $A$:
\begin{eqnarray} \label{equ:V_final}
V_{mn} &=& \frac{{\rm cov}(O^A_{m, 10}, O^A_{n,10}) + {\rm cov} (O^B_{m,
    10}, O^B_{n,10})}{2} + \frac{{\rm cov}(O^A_{m, 100}, O^A_{n,100})
  + {\rm cov} (O^B_{m, 100}, O^B_{n,100})}{2} \nonumber \\
 &+& \delta_{mn}\left( k_m^{({\rm syst})} \, \frac{O^A_{m, 10} + O^B_{m,
      10}}{2} \right)^2 \, . 
\end{eqnarray}
Here the subscripts $10$ and $100$ refer to data sets corresponding to
integrated luminosities of $\unit[10]{fb^{-1}}$ for the measurement and
$\unit[100]{fb^{-1}}$ for the prediction.\footnote{Recall that we are
  predicting the expectation values of our observables. Most of these
  expectation values are independent of the integrated luminosity. The
  only exception is the total number of events after cuts, where the
  prediction obviously is one tenth of the number of accepted events
  in the larger data sample.} The first term in eq.(\ref{equ:V_final})
therefore describes the (expected) statistical uncertainties
associated with the measurements, while the second term comes from the
uncertainty on the predictions due to finite (although large) Monte
Carlo statistics. Finally, the last term describes systematic
uncertainties, which we assume to contribute only to the diagonal
elements of the covariance matrix. We assume that the systematic
uncertainty for a given observable $O_m$ is some fixed fraction
$k_m^{({\rm syst})}$ of this observable. This is the usual ansatz at
least for systematic theoretical uncertainties; experimental
systematic errors also tend to decrease with increasing luminosity.
Our numerical choices for the $k_m^{({\rm syst})}$ will be discussed
below. 

When calculating the total $\chi^2_{AB}$ from eq.(\ref{equ:chi2AB}),
and the corresponding $p-$value from eqs.(\ref{equ:p}) and
(\ref{equ:f}), we again only include relevant observables, defined
exactly as in Subsection~\ref{sec:teststat}. For the 283 pairs found
indistinguishable in ref.\cite{bib:inverseProblem}, we find that on
average 32 observables are included in the comparison. Recall that
about 40 observables were included in the self--comparison described in the
Subsection~\ref{sec:teststat}. This is expected, since for most
observables {\em both} data sets have to contain a minimal number of
events of a given type for a certain observable to be included, as
detailed in Table~\ref{tab:nmin}. If two different parameter sets are
compared it is more likely that one of the sets does not satisfy this
criterion, thereby reducing the number of observables we use. 

On average about $25,000$ events pass all our cuts, for a cut
efficiency of about $\unit[30]{\%}$. In most cases a large majority of these
events again contains no or only one isolated charged lepton (electron
or muon), i.e. the bulk of events belongs to the first three classes
described in Subsection~\ref{sec:observables}.

In the following three Subsections we present results for four
different assumptions regarding systematic errors and Standard Model
backgrounds. 

\subsection{No Systematic Errors, no Backgrounds}

Our first result is that in an ideal world, where the Standard
Model background as well as all systematic errors can be neglected,
{\em all} of the 283 degenerate pairs are distinguishable, i.e. lead
to a $p-$value below $0.05$; the mean $p-$value of these 283 pairs is
$6.8 \cdot 10^{-5}$. Recall that ref.\cite{bib:inverseProblem} also
uses estimated $\unit[95]{\%}$ c.l. intervals to define distinguishability.

\subsection{With Systematic Errors, no Backgrounds}
\label{sec:delP}

Arkani--Hamed {\it et al.} did include systematic errors. In order to
allow for a direct comparison with their results, we also assume a
relatively large systematic uncertainty of $\unit[15]{\%}$ on the total number
of events after cuts, i.e. $k^{({\rm syst})}_N = 0.15$;
this includes the luminosity error on the experimental side, and the
uncertainty in the prediction of the total cross section on the theory
side. Again following ref.\cite{bib:inverseProblem} we assign much
smaller systematic errors of $\unit[1]{\%}$, i.e. $k_m^{({\rm syst})} = 0.01$,
to all other observables. This inclusion of systematic uncertainties
does not change the criteria defining which observables are included
in the comparison. After these systematic uncertainties are included,
260 of the pairs deemed indistinguishable by Arkani--Hamed {\it et
  al.} still have $p-$values below 0.05, i.e. can be distinguished. We
find that only 23 pairs have $p > 0.05$, which increases the mean
$p-$value to $0.038$. Note that most pairs have very small
$p-$values. For example, only 51 pairs have $p > 10^{-4}$. 

In order to interpret the distribution of $p-$values we consider the
quantity $(\Delta P_{AB})^2$ introduced by Arkani--Hamed {\it et al.},
which is defined by
\begin{equation} \label{equ:dP2}
(\Delta P_{AB})^2 = \frac{1}{n_{para}} \sum\limits_{i = 1}^{n_{para}}
\left( \frac{P^A_i - P^B_i}{\bar{P}^{AB}_i} \right)^2 \, .
\end{equation}
Here $P^A_i$ and $P^B_i$ are the values of the $i-$th parameter for
parameter sets $A$ and $B$, respectively, and $n_{para} \leq 15$ is
the number of parameters used in a given comparison. $\bar{P}^{AB}_i =
(P^A_i+P^B_i)/2$ is the average value of the $i-$th parameter for both
sets.  Since all our parameters are positive, $(\Delta P_{AB})^2$ can
take values in the range $[0, \, 4]$. For small values it corresponds
approximately to the relative difference squared, i.e. $ (\Delta
P_{AB})^2 = 0.01$ implies a $\unit[10]{\%}$ difference between the
parameters of the parameter sets $A$ and $B$.

\begin{figure*}[t]
\centering	
\begin{minipage}[b]{8cm}
\includegraphics[width=8cm]{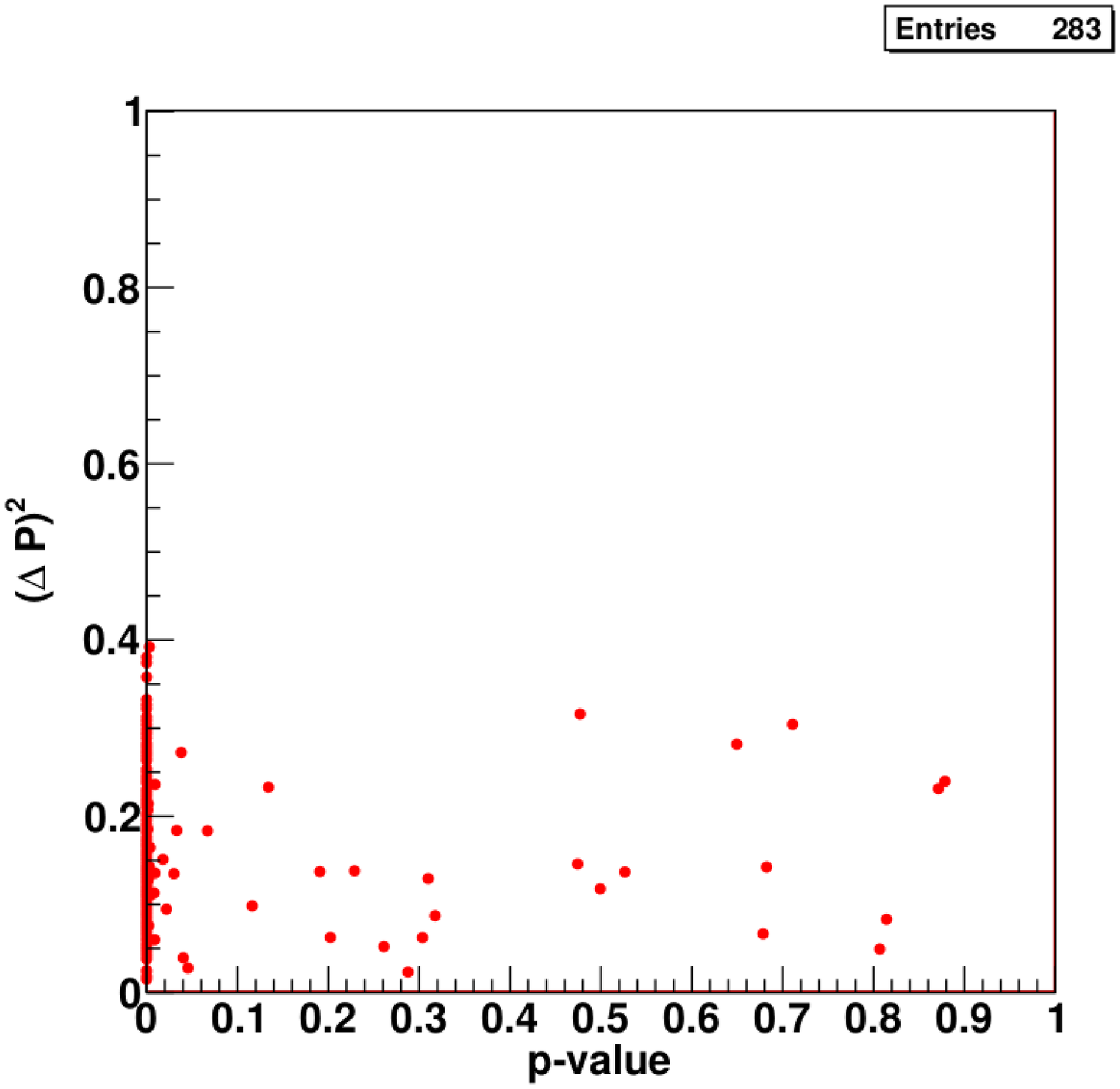}
\end{minipage}
\begin{minipage}[b]{8cm}
\includegraphics[width=8cm]{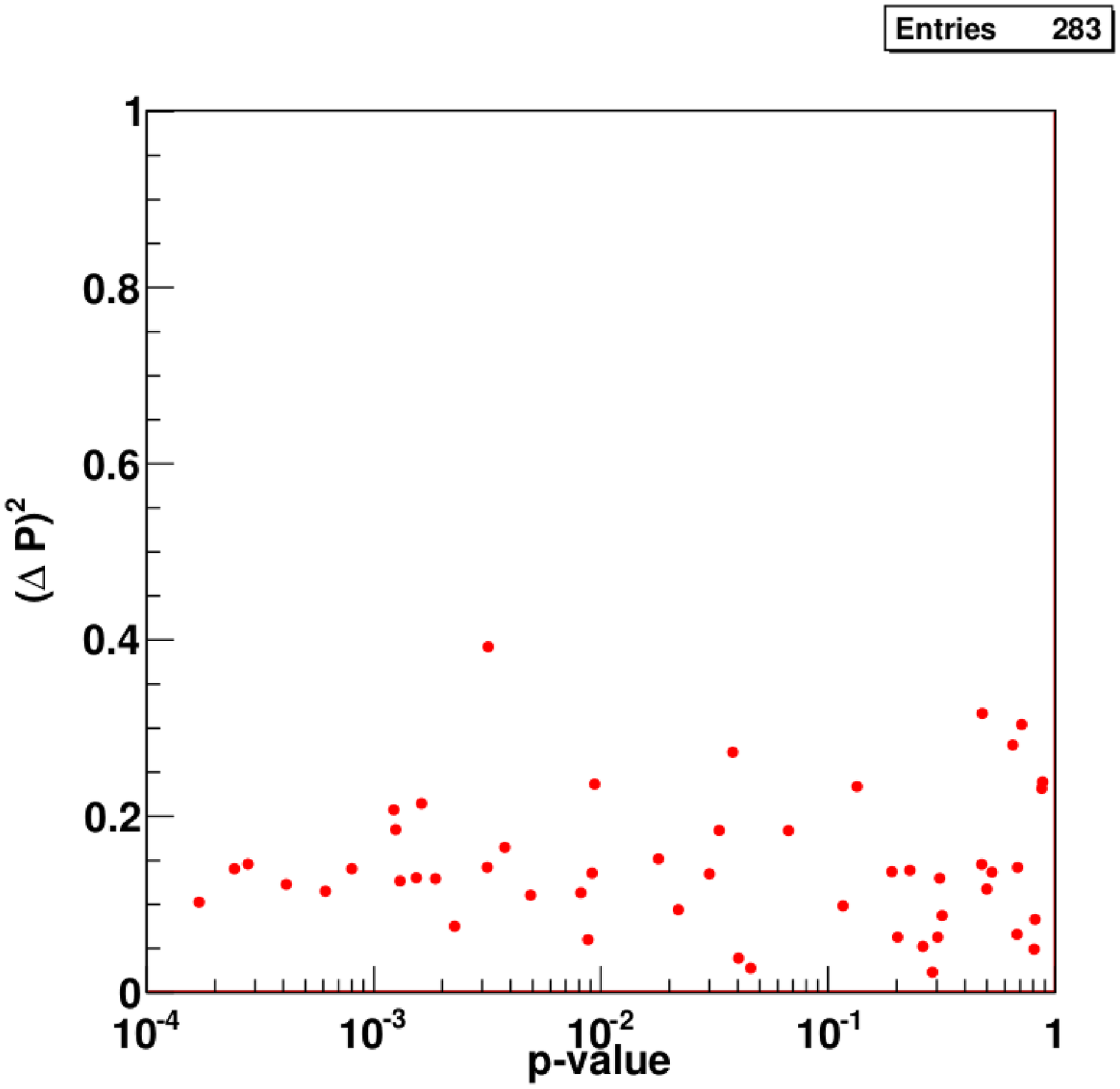}
\end{minipage}
\caption{The total parameter difference $(\Delta P)^2$ defined in
  eq.(\ref{equ:dP2}) including all 15 parameters versus the $p-$value
  for the 283 pairs deemed indistinguishable in
  ref.\cite{bib:inverseProblem}, including systematic errors. The
  $p-$value is shown on a linear (left) and logarithmic scale (right);
  in the latter case pairs with with $p < 10^{-4}$ are not shown.}
\label{fig:DeltaPtot}
\end{figure*}

Clearly in the limit $(\Delta P_{AB})^2 \rightarrow 0$ the two
parameter sets $A$ and $B$ become identical, and hence
indistinguishable. One could naively expect that pairs of parameter
sets giving a large value of $(\Delta P_{AB})^2$ should be relatively
easy to distinguish. However, in Fig.~\ref{fig:DeltaPtot} we see that
is not necessarily the case: there are pairs of parameter sets where
{\em both} $(\Delta P_{AB})^2$ and the $p-$value are quite large. This
indicates that, at least under certain circumstances, the observables
we are considering are not sensitive to some of our parameters. We
remind the readers that all our parameter sets have relatively light
superparticles, not much above the TeV scale, so the total rate of
SUSY events is significant. Recall also that {\em all} pairs of
parameter sets we are considering here were considered
indistinguishable by Arkani--Hamed et al., including parameter sets
with relatively large $(\Delta P_{AB})^2$. As noted above, our
analysis shows that most of these pairs can in fact be distinguished.

In the right frame of Fig.~\ref{fig:DeltaPtot}, which uses a logarithmic scale for the $p-$value,
the lower edge of the populated region includes smaller $(\Delta
P_{AB})^2$ for larger $p-$values, which conforms with naive
expectations. However, in the left frame, which uses a linear $x-$axis
and includes all 283 pairs, we also find some pairs where both
$(P_{AB})^2$ and $p$ are quite small. This can e.g. happen if a small
variation of some mass opens or closes some decay channel
characterized by large couplings and hence potentially large branching
ratios, which will significantly alter the final state of many SUSY
events. The main conclusion therefore is that there is relatively
little correlation between the $p-$value and the average parameter
difference $(\Delta P_{AB})^2$ computed from all 15 input parameters.

\begin{figure*}[t!]
\centering	
\begin{minipage}[b]{8cm}
\includegraphics[width=8cm]{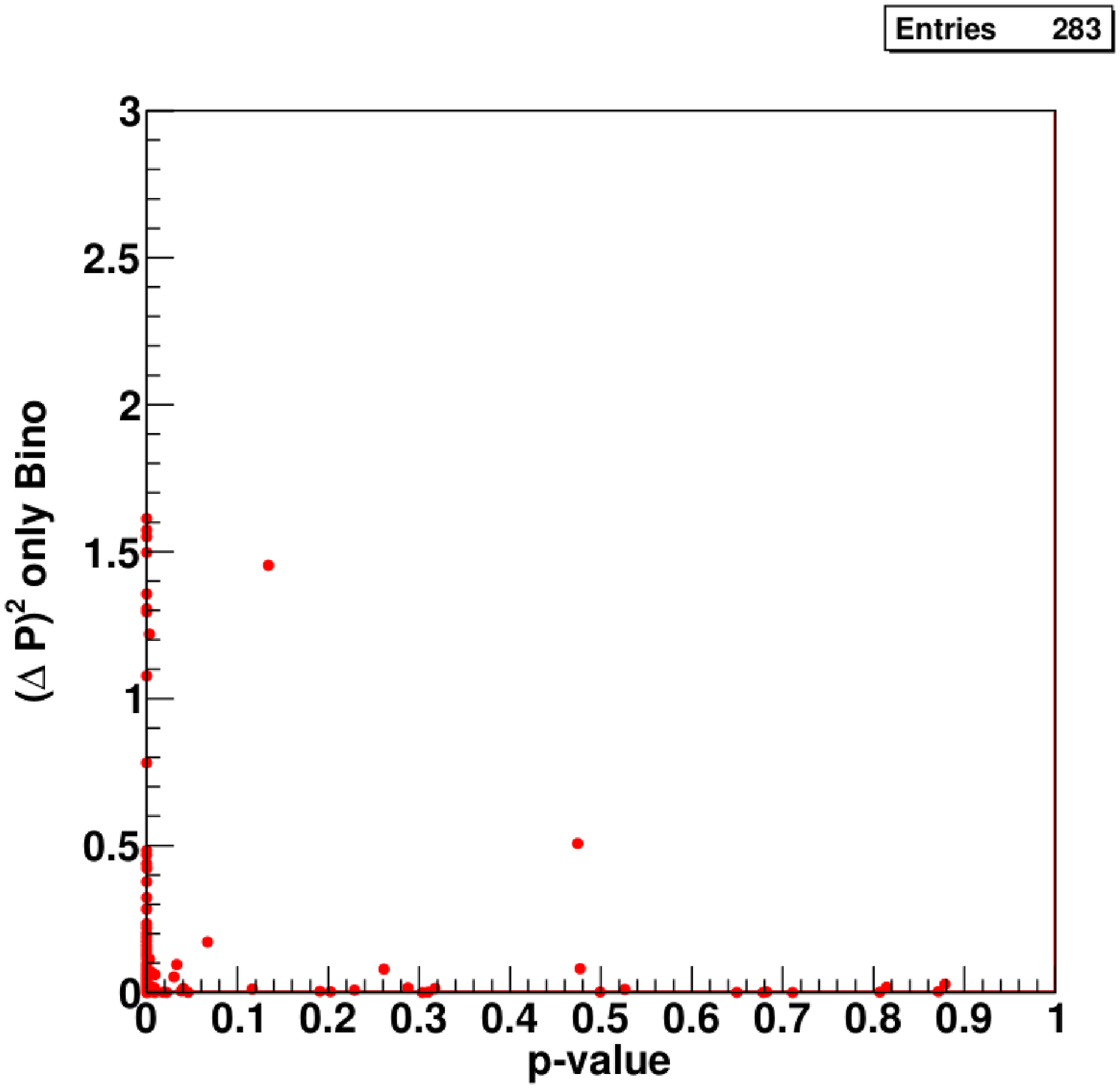}
\end{minipage}
\begin{minipage}[b]{8cm}
\includegraphics[width=8cm]{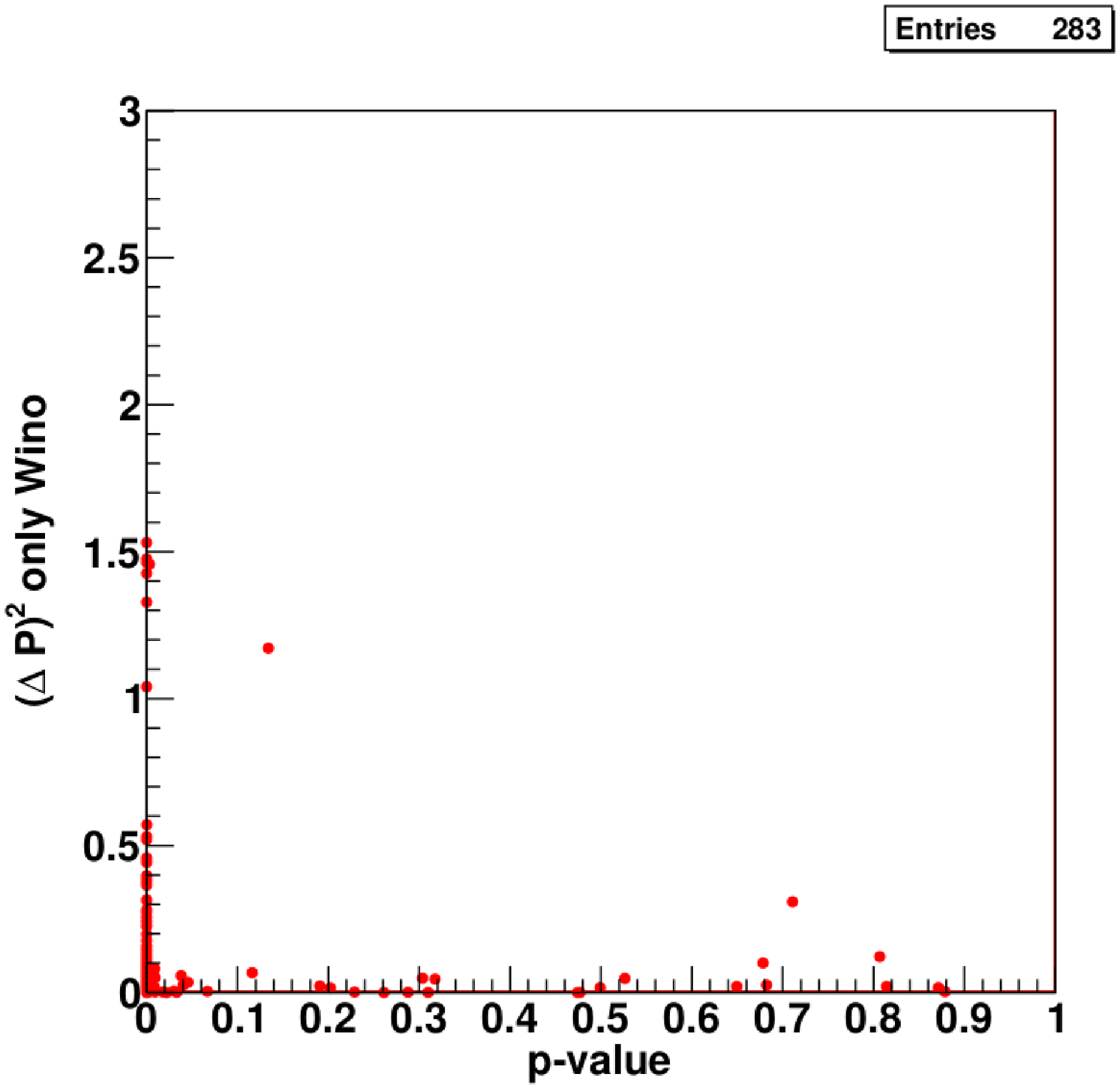}
\end{minipage}
\begin{minipage}[b]{8cm}
\includegraphics[width=8cm]{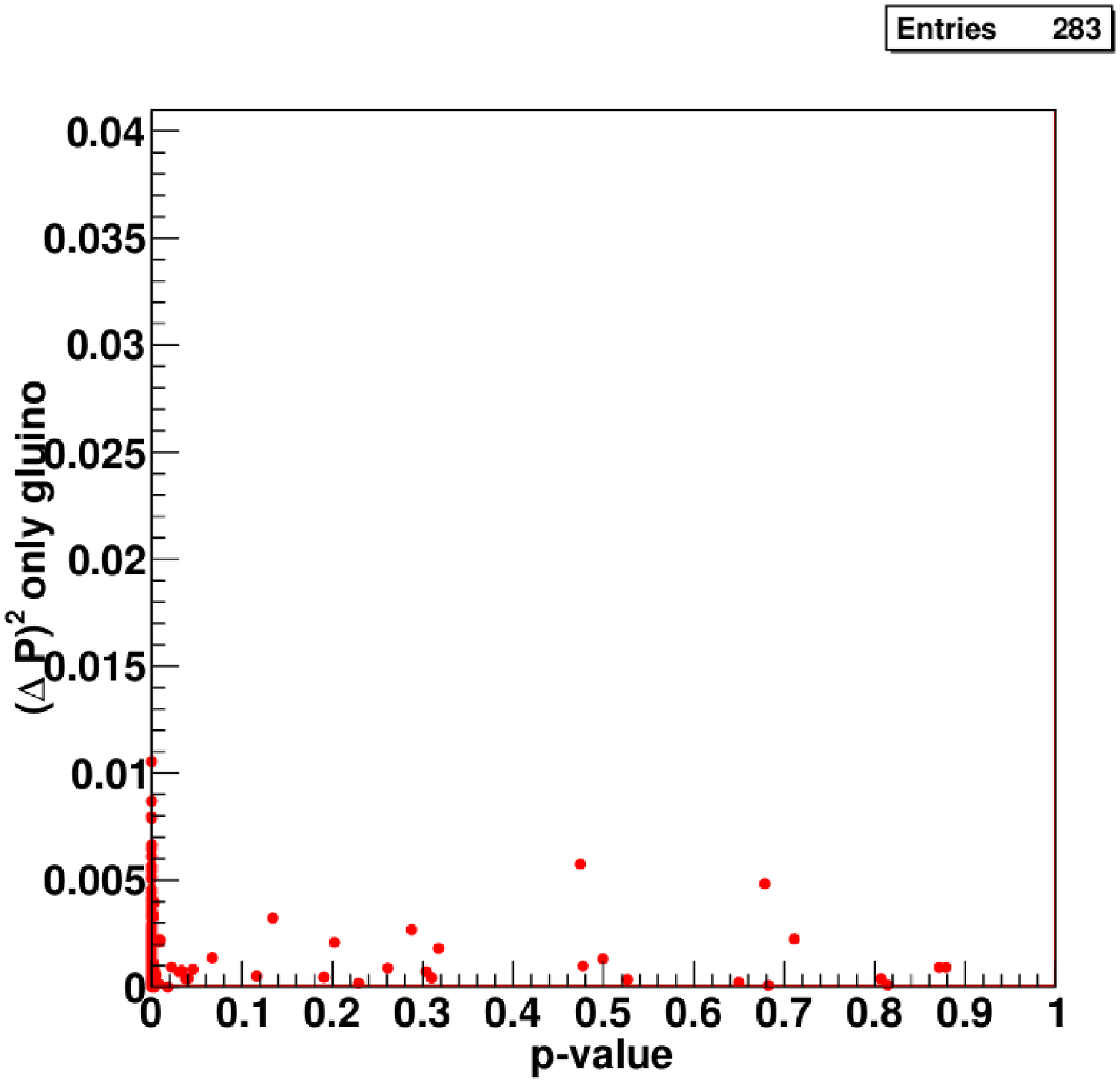}
\end{minipage}
\begin{minipage}[b]{8cm}
\includegraphics[width=8cm]{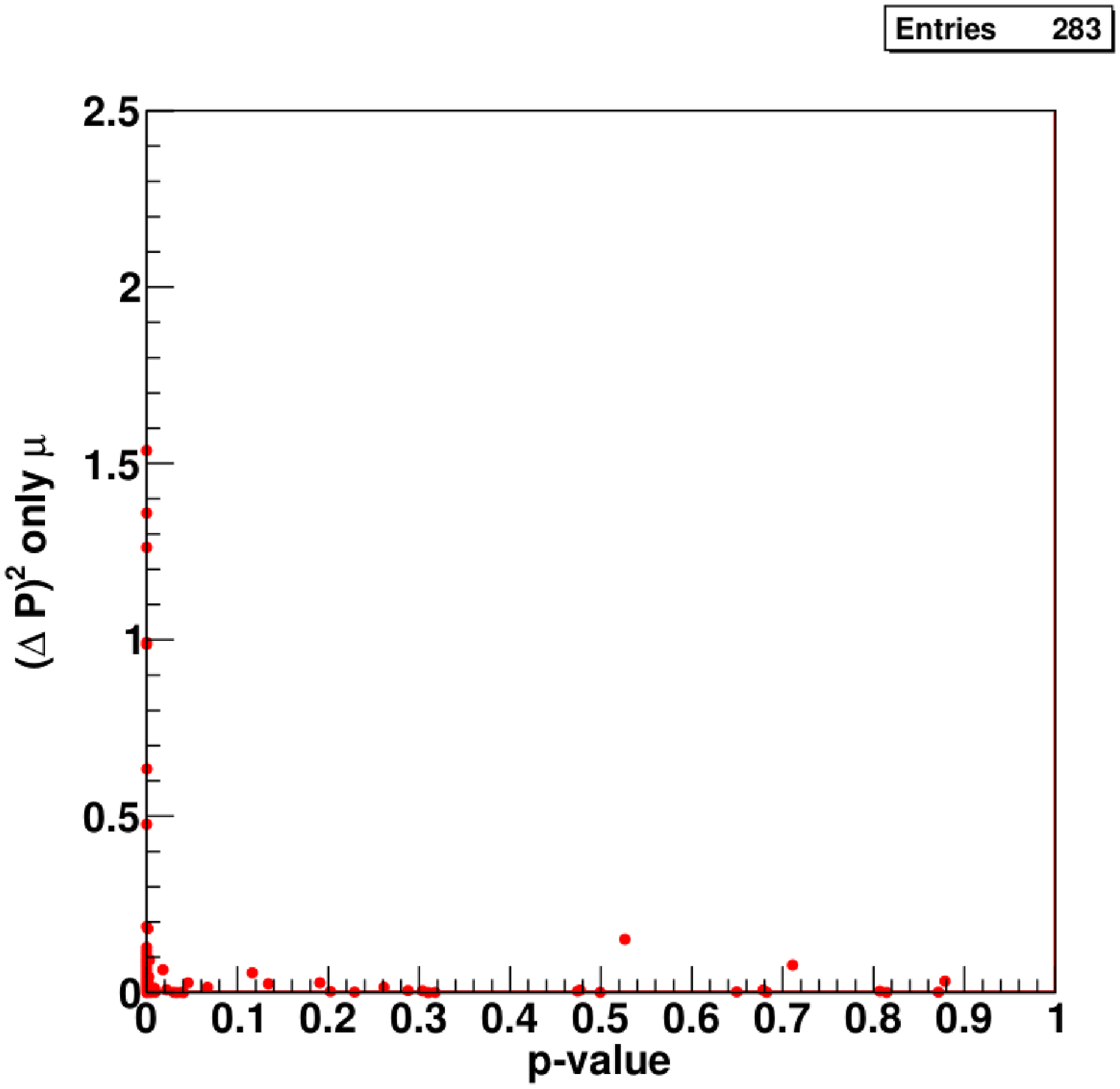}
\end{minipage}
\caption{As in Fig.~\ref{fig:DeltaPtot}, except that only a single
  parameter has been used in the calculation of $(\Delta P)^2$: $M_1$
  (top left), $M_2$ (top right), $M_3$ (bottom left), $\mu$
  (bottom right).} 
\label{fig:DeltaPino}
\end{figure*}

In order to gain a better understanding of this result, we consider
results where only a subset of our 15 free parameters has been used in
the calculation of $(\Delta P_{AB})^2$. We begin with
Fig.~\ref{fig:DeltaPino}, which shows results where $(\Delta
P_{AB})^2$ has been calculated from a single parameter in the
gaugino--higgsino sector. We see that pairs of parameter sets that
have $p > 0.05$ have gluino masses that differ by at most
$\unit[7]{\%}$. This is not surprising since, given our assumption $|M_3|
\lsim \unit[1]{TeV}$, a significant change of the gluino mass inevitably leads
to a significant change of the total number of SUSY events.

Among the parameters in the electroweak gaugino--higgsino sector,
$\mu$ shows the smallest difference between pairs of parameter sets
with $p > 0.05$, with $(\delta \mu_{AB})^2 \lsim 0.15$; this is to be
contrasted with $(\delta \mu_{AB})^2 \lsim 1.5$ if all pairs
considered indistinguishable in ref.\cite{bib:inverseProblem} are
considered. The fact that $\mu$ shows the smallest difference in pairs
that are difficult to distinguish can be understood from the
observation that this parameter not only largely determines the masses
of two neutralinos and one chargino, but also affects mixing among
third generation sfermions. A change of $\mu$ therefore generally
changes more observables than changes of the electroweak gaugino mass
parameters $M_1$ and $M_2$ do.

Most pairs of parameter sets with $p > 0.05$ also have quite similar
values of these latter parameters, but there are a few outliers. This
is due to the fact that our counting observables are mostly sensitive
to the ordering of the three electroweak gaugino--higgsino mass
parameters. For example, reducing the smallest of these parameters
even more may have little impact on branching ratios, and may not
change our sole kinematic quantity $H_T$ very much as long as the
lightest neutralino remains sufficiently light. Conversely, if the
bino mass is larger than the gluino mass and also larger than the
masses of the sleptons, few bino--like neutralinos will be produced at
the LHC, so that $|M_1|$ cannot be determined very well by LHC
experiments.

The two points with $p \approx 0.13$ and $(\Delta P_{AB})^2 > 1$ in the two
upper frames of Fig.~\ref{fig:DeltaPino} correspond to the same pair
of parameters. Here the values of $M_1$ and $M_2$ are (approximately)
interchanged between the two sets. The existence of such ``mirror
pairs'' has also been noticed in ref.\cite{bib:inverseProblem}. Note
that there is a strong hierarchy between these two parameters in both
cases; moreover, $\mu$ is quite large. As a result, basically only the
lightest neutralino is produced in LHC events, irrespective of the
ordering of $M_1$ and $M_2$. In the scenario with $M_2 \ll M_1$, the
lightest chargino is also produced frequently; however, since its mass
splitting to the lightest neutralino is only ${\cal O}(100)$ MeV, the
lighter chargino effectively behaves like the lightest neutralino in
this scenario, as far as our observables are concerned.

\begin{figure*}[t]
\centering	
\begin{minipage}[b]{8cm}
\includegraphics[width=8cm]{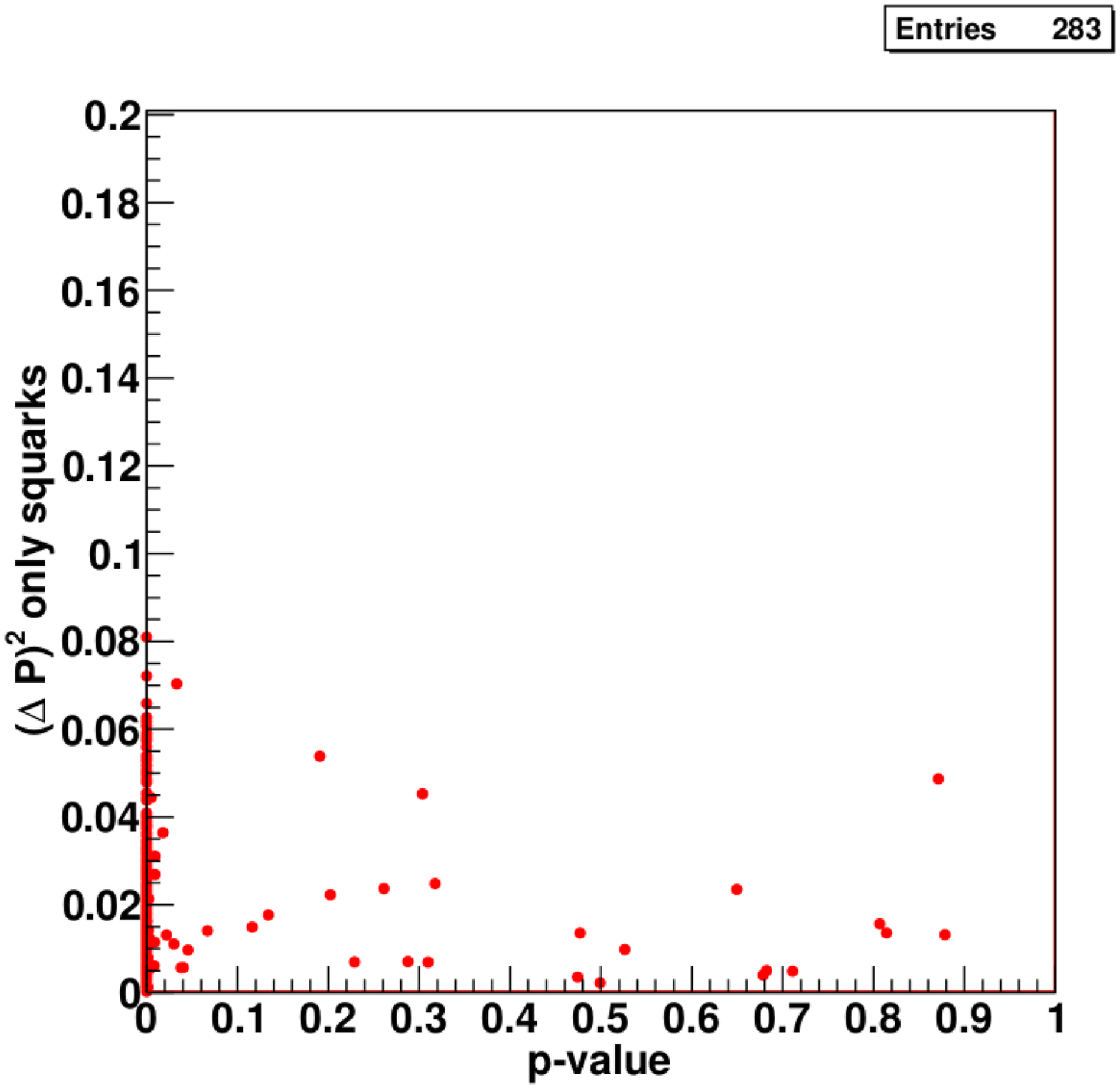}
\end{minipage}
\begin{minipage}[b]{8cm}
\includegraphics[width=8cm]{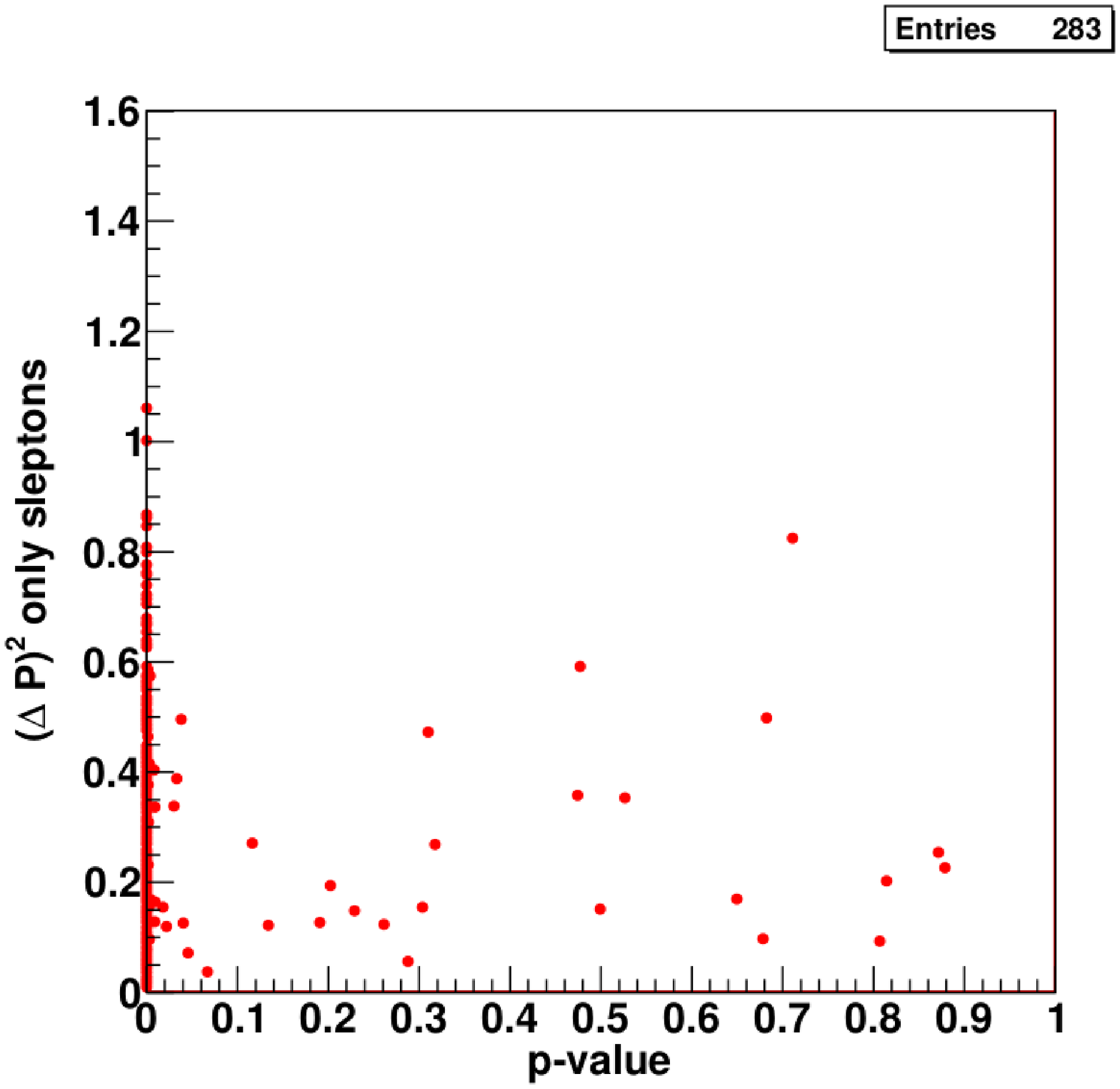}
\end{minipage}
\caption{As in Fig.~\ref{fig:DeltaPtot}, except that the calculation
  of $(\Delta P)^2$ includes only the squark (left) or the slepton mass
  parameters (right).}
\label{fig:DeltaPsfermion}
\end{figure*}

Figure~\ref{fig:DeltaPsfermion} shows results for sfermion mass
parameters. In the left frame the sum in eq.(\ref{equ:dP2}) runs over
the six squark mass parameters of our model. We see that the squark
masses differ relatively little within pairs of parameter sets with $p
> 0.05$, although the spread is significantly larger than for the
gluino mass (up to about $\unit[24]{\%}$ as compared to up to about
$\unit[7]{\%}$). The reason is that the cross section for producing a single
type of squark is much smaller than the cross section for gluino pair
production, for equal masses; this is true in particular for second
and third generation squarks, for which no $t-$ or $u-$channel
processes involving two valence ($u$ or $d$) quarks in the initial
state are available. Moreover, one can again set up ``mirror pairs'',
e.g. by swapping the masses of $SU(2)$ doublet ($L-$type) and singlet
($R-$type) squarks if all squarks decay directly into the lightest
neutralino.

The right frame of Fig.~\ref{fig:DeltaPsfermion} shows corresponding
results for the slepton mass parameters. Evidently parameter pairs
with $p > 0.05$ can have quite different slepton masses. The
observables we are using are sensitive to slepton masses essentially
only if sleptons are produced in the decays of strongly interacting
sparticles. While the conditions (\ref{bound}) make it likely that
such decays are kinematically possible for some strongly interacting
sparticle(s), they by no means ensure that the corresponding cross
section times branching ratio is sizable. In particular, we saw above
that bino--like neutralinos may be produced very rarely at the LHC. If
in addition the wino--like states are lighter than the sleptons, very
few sleptons will be produced in the decays of gluinos and squarks,
and our observables will not be sensitive to slepton masses.

\begin{figure*}[t]
\centering
\resizebox{0.5\textwidth}{!}{
\includegraphics{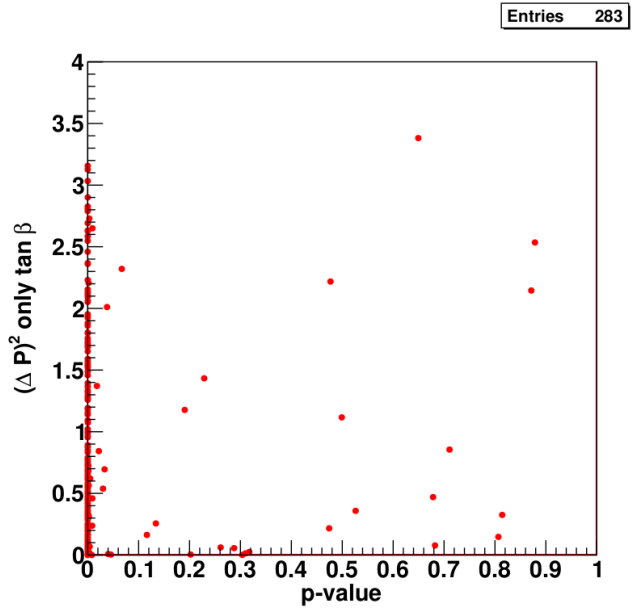}}
\caption{As in Fig.~\ref{fig:DeltaPtot}, except that $(\Delta P)^2$ is
  calculated only from $\tan \beta$.}  
\label{fig:DeltaPtanb}
\end{figure*}

Finally, Fig.~\ref{fig:DeltaPtanb} shows that $\tan\beta$ has the
largest spread among all our parameters, even when only pairs of
parameter sets with $p > 0.05$ are considered. In principle this
parameter has influence on many observables, since it appears in the
chargino and neutralino mass matrices and also affects all sfermion
masses. However, in most cases the dependence of the physical masses
and couplings on this parameter is weak. It can be enhanced if two
charginos or two neutralinos are close in mass, or if $\tan\beta$ is
very large, in which case it modifies $\tilde b$ and $\tilde \tau$
mixing significantly.

\subsection{Including Standard Model Background}

We next include Standard Model backgrounds, which had been ignored in
ref.~\cite{bib:inverseProblem}. We include backgrounds from $Z + {\rm jets}$, $W + {\rm jets}$ and $t \bar{t}$ as well as single top production; the
latter is generated using the programs MadGraph 5 \cite{bib:madgraph}
and Herwig++, and the other backgrounds are simulated directly using
Herwig++. For the first two backgrounds, only leptonic $W$ and $Z$
decays are included, which have a chance to pass our cut on the
missing transverse momentum.\footnote{In order to reduce the number of
  generated events we require a minimum transverse momentum of the $Z$
  and $W$, respectively, of $\unit[100]{GeV}$ at the parton level,
  i.e. prior to showering. This has little influence on the total
  number of events which pass the cuts because events with lower
  parton--level transverse momentum have almost no chance to pass our
  missing $p_T$ cut.} All background (and signal) contributions are
only considered to leading order in perturbation theory.

Altogether we find $29,052$ background events after cuts for a data
sample of $\unit[10]{fb^{-1}}$, yielding a signal to background ratio
of very roughly 1 to 1. We add these to all our ``measurement''
samples. An independent background simulation for
$\unit[100]{fb^{-1}}$ of data yielded $293,875$ events after cuts,
which we add to our ``prediction'' samples. The biggest source of
background is top pair production, followed by the production of a $W$
or $Z$ boson in association with several jets. Of course, the
background is also included in the calculation of the covariance
matrix.

Recall that a given observable is only included in the calculation of
the overall $\chi^2_{AB}$ if the corresponding event class contains a
minimal number of signal events, as listed in Table~\ref{tab:nmin}. We
now demand instead that there should be a SUSY signal of at least $3
\, \sigma$ statistical significance for an observable to be considered
relevant, i.e. we only include observables if the number of
supersymmetric events in the corresponding class is greater or equal
three times the square--root of the number of background
events.\footnote{The analogous requirement on our first observable,
  the total number of events after cuts, is satisfied for all
  scenarios we are considering.} In case of the event fractions in a
given event class $O_{1,c}$, this requirement has to be satisfied for
at least one of the two pairs of parameters in a given comparison; all
other observables are included only if this new requirement is
satisfied for both pairs of parameters.

Including these backgrounds, but ignoring all systematic errors leads
to only one parameter pair with $p > 0.05$. However, including
systematic errors as well as SM backgrounds increases the number of
pairs that cannot be distinguished ``at $\unit[95]{\%}$ confidence level'' to
46. This means that our simple algorithm can resolve more than
$\unit[80]{\%}$ of the ``degenerate pairs'' found by Arkani--Hamed et al.,
even after SM backgrounds are included.

\section{Summary and Conclusions}
\label{sec:SummaryandConclusions}

In this paper we investigated the question to which extent a single
$\chi^2$ variable can be used to discriminate between different
parameter sets in a general MSSM with 15 free parameters. This
analysis was triggered by ref.\cite{bib:inverseProblem}, which
identified 283 pairs of parameter sets that were claimed to be
indistinguishable by LHC experiments with $\unit[10]{fb^{-1}}$ of data taken
at $\sqrt{s} = \unit[14]{TeV}$. In our analysis we used far fewer observables
(up to 84, as compared to up to 1808), but took care to properly
include all correlations, so that our $\chi^2_{AB}$ defined in
eq.(\ref{equ:chi2AB}) should behave like a proper $\chi^2$
distribution in the limit of Gaussian statistics; we checked this
explicitly in Sec.~\ref{sec:teststat}. We saw in the Introduction that
this does not seem to be true for the ``$\chi^2-$like'' variable
considered in ref.~\cite{bib:inverseProblem}. We also improved the
analysis by including initial state showering and the underlying event
when simulating our signal events. We also analyzed the effects of
Standard Model backgrounds.

\begin{table*}
\centering
\caption{Number of pairs of parameter sets with $p > 0.05$, out of the
  283 pairs deemed indistinguishable in
  ref.~\cite{bib:inverseProblem}, for different levels of
  sophistication of our analysis (with or without systematic errors
  and Standard Model backgrounds). The mean and median values of $p$
  for all 283 pairs are also given.} 
\vspace*{3mm}
\begin{tabular}{|c|c||c|c|c|}
\hline
Syst. Errors & Backgrounds & no. of pairs with $p>0.05$ & $\bar p$ &
median $p$ \\
\hline
No & No & 0 & $6.8 \cdot 10^{-5}$ & $3.6 \cdot 10^{-146}$ \\
Yes & No & 23 & $0.038$ & $1.1 \cdot 10^{-36}$ \\
\hline
No & Yes & 1 & $0.0030$ & $2.6 \cdot 10^{-79}$ \\
Yes & Yes & 46 & $0.079$ & $1.4 \cdot 10^{-13}$ \\
\hline
\end{tabular}
\label{tab:Results}
\end{table*}

Our results are summarized in Table~\ref{tab:Results}. We see that
under the conditions of ref.\cite{bib:inverseProblem} (with systematic
errors but without backgrounds) all but 23 of the supposedly 283
indistinguishable pairs of parameters can in fact be distinguished at
the $\unit[95]{\%}$ confidence level. The median $p-$value for all 283 pairs
is then of order $10^{-36}$. Including also SM backgrounds increases
the number of pairs with $p > 0.05$ to 46, and the median $p-$value
increases by a factor $10^{23}$ -- but it still remains very small, of
order $10^{-13}$. We therefore conclude that our analysis strategy can
greatly alleviate the ``inverse problem'' at the LHC.

This Table also indicates that the inclusion of systematic uncertainties
has a bigger impact than the inclusion of SM backgrounds. This is
somewhat worrisome, since our ansatz for the systematic errors is
quite ad hoc. Experimental systematic errors can only be estimated
properly by our experimental colleagues. The rather small total
systematic errors we include can only be realized once quantum
corrections to all relevant cross sections and branching ratios have
been computed. We are confident that this will happen if and when
superparticles are discovered; in this we are encouraged by a very
recent NNLO calculation of the cross section for $t \bar t$ production
from $q \bar q$ annihilation \cite{bib:nnlo}, which reduces the higher
order uncertainty, estimated by varying unphysical factorization and
renormalization scales, to a value below $\unit[3]{\%}$. Recall also that
all sparticles have masses at or below $\unit[1]{TeV}$ in the parameter sets
considered in ref.\cite{bib:inverseProblem}, leading to quite large
event samples at the LHC; this reduces the importance of
backgrounds. As noted earlier, many of these spectra are most likely
excluded by existing LHC searches; we nevertheless used them in our
analysis to ensure that our results can be compared directly with
those of ref.\cite{bib:inverseProblem}.

While we consider our results to be quite encouraging, we saw in
Sec.~\ref{sec:delP} that in some cases our observables cannot
discriminate between sets where at least some of the input parameters
differ quite significantly, even if backgrounds are ignored. The pairs
with $p > 0.05$ and the largest overall difference between input
parameters, measured via the quantity $(\Delta P_{AB})^2$ introduced
in eq.(\ref{equ:dP2}), all have squark and gluino masses near the
upper end of the scanned range, i.e. relatively low sparticle
production cross sections and hence relatively small event
samples. Moreover, they all have wino mass $M_2 \ll |\mu|, \, |M_1|$,
i.e. a light wino--like lightest chargino with small mass splitting to
the neutralino. In fact, the mass splitting is $\lsim 200$ MeV, giving
macroscopic decay length for the lightest chargino \cite{bib:cdg}. A
measurement of this decay length would give an additional constraint
on our input parameters. For example, it would break the degeneracy
between pairs that basically only differ by $M_1 \leftrightarrow M_2$;
see the discussion of Fig.~\ref{fig:DeltaPino}. 

Similarly, we have not tried to isolate certain supersymmetric
production channels. Our cuts have been devised solely with the
purpose of suppressing SM backgrounds; no attempt has been made to
suppress ``supersymmetric backgrounds'' to certain channels. For
example, direct slepton pair production might be detectable at the LHC
under certain circumstances \cite{bib:slep_at_lhc}. This should help
to resolve at least some of our degeneracies.

Note also that our simulation does not include mixed ${\cal
  O}(\alpha_S \alpha_W)$ contributions to squark production cross
sections, which are mostly due to interference between QCD diagrams
and diagrams where an electroweak gaugino is exchanged in the $t-$ or
$u-$channel. These contributions can change some squark pair cross
sections by tens of percent \cite{bib:bddk1}, thereby offering another
handle on the electroweak gaugino masses; they can also help to
discriminate between $SU(2)$ singlet and doublet squarks, which can be
difficult if only a single neutralino state is accessible to squark
and gluino decays.

Finally, we have not seriously attempted to optimize the cuts, or the
selection of observables, with a view of improve the
distinguishability between pairs of parameter sets. We feel that this
can be done in a meaningful manner only if many more than the 283
pairs of parameter sets we are studying are analyzed, which
unfortunately is beyond the power of our computer resources. Besides,
given that the LHC is running and producing data, the usefulness of
vast scans of parameter space for the purpose of optimizing methods to
distinguish between discrete sets of input parameters appears somewhat
questionable. 

Instead the analysis presented here had two objectives. First, we
wanted to alleviate concerns raised in ref.\cite{bib:inverseProblem}
that LHC experiments may not be able to determine many SUSY parameters
even if sparticles are quite light. As argued above, this goal was
largely achieved. This indicates that our observables should also be
useful for determining the values of supersymmetric parameters, rather
than ``only'' for the purpose of discriminating between discrete sets
of input parameters. Our second goal is therefore to motivate further
studies, where (some of) our observables are used to determine
supersymmetric parameters, as alternative, or in addition, to the
methods that have been used so far.

\subsection*{Acknowledgments}
We thank Jesse Thaler for providing the parameter sets analyzed by
us. NB wants to thank the ``Bonn-Cologne Graduate School of Physics
and Astronomy'' and the ``Universit\"at Bonn'' for the financial
support. This work was partially supported by the BMBF--Theorieverbund
and by the Helmholtz Alliance ``Physics at the Terascale''.

\begin{appendix}
\section{Definition of Observables}
\label{sec:leptonsJetsAndHT}

In this Appendix we describe in detail how our observables are
defined.  In particular, charged leptons and jets have to fulfill
certain acceptance cuts to be counted. The $\tau-$jets which arise
from hadronically decaying $\tau-$leptons are in general just called
taus here. The determination of $b-$ and non$-b-$jets is also
explained in the following. The final subsections deal with the
calculation of the missing transverse momentum $\cancel{p}_T$ and the
observable $H_T$. 

Note that we count visible particles as measurable only if they have
pseudorapidity $|\eta| < 5$; neutrinos and the lightest neutralino are
not considered visible, since they (usually) do not interact in the
detector.

\subsection{Electrons and Muons}
\label{sec:emu}

We only consider {\em isolated} electrons and muons, i.e. a charged
lepton $l$ has to fulfill the following criteria:
\begin{itemize}
\item $ |\eta^l| < 2.5 $
\item $ p^l_T > \unit[10]{GeV} $
\item For all measurable particles with $\Delta R = \sqrt{(\eta^l -
    \eta)^2 + (\phi^l - \phi)^2} \leq 0.2$ the transverse energy $E_T$
  is summed. This sum has to be $\sum E_T < \unit[5]{GeV}$  
\end{itemize}

\subsection{Taus}
\label{sec:tau}

Only hadronically decaying taus, i.e. $\tau-$jets, are considered
here; the electrons and muons produced in leptonic $\tau$ decays are
counted as all other charged leptons, if they satisfy the criteria
listed in the previous Subsection. We use generator information to
check whether a given $\tau-$lepton indeed decays hadronically. The 
four--momentum $p^{\tau}$ of the $\tau-$jet is then defined as
the difference between the four--momentum of the parent $\tau-$lepton
and the four--momentum of the corresponding $\nu_\tau$. We impose the
following acceptance cuts on the $\tau-$jets:
\begin{itemize}
\item $ |\eta^{\tau}| < 2.5 $
\item $ p^{\tau}_T > \unit[20]{GeV} $
\end{itemize}
Furthermore the $\tau-$jet should be isolated, i.e. there must not be
any charged hadrons with $p_T > \unit[1]{GeV}$ and no photons with
$p_T > \unit[1.5]{GeV}$ within a cone $\Delta R = \sqrt{(\eta^{\tau} -
  \eta)^2 + (\phi^{\tau} - \phi)^2} < 0.5$ around the tau; note that
this condition is usually more stringent than that used for electrons
and muons. Moreover, a $\tau-$jet satisfying all conditions is only
tagged with a $\unit[50]{\%}$ probability. Recall also that some
$\unit[35]{\%}$ of all $\tau-$leptons decay leptonically. Altogether
we thus see that a sample of events containing equal numbers of
electrons, muons and $\tau-$leptons with identical kinematical
distributions would indeed yield equal numbers of observed electrons
and muons in our simulation, but far fewer identified $\tau-$leptons. Note finally that identified $\tau-$jets are not included in
the identification of other jets.

\subsection{Jets}
\label{sec:jets}

Jets are determined with the program FastJet \cite{bib:fastjet}
using the anti$-k_t$ algorithm. All measurable particles are taken
into account, except for isolated electrons and muons that satisfy the
criteria of Subsec.~\ref{sec:emu} as well as isolated $\tau-$jets
defined in Subsec.~\ref{sec:tau}. We use the following parameter
choices for the jet finding algorithm:
\begin{itemize}
\item double dou\_R = 0.5
\item fastjet::RecombinationScheme RecSch\_scheme = fastjet::E\_scheme
\item fastjet::Strategy Stra\_strategy = fastjet::Best
\item [$ \rightarrow $] fastjet::JetDefinition
  JetDef\_def(JetAlgo\_algo, dou\_R, RecSch\_scheme, Stra\_strategy) 
\end{itemize}
Reconstructed jets are only counted if they satisfy:
\begin{itemize}
\item $ p^{j}_T > \unit[20]{GeV} $
\item $ |\eta^{j}| < 4.8 $
\end{itemize}

In order to determine whether a given jet originates from a $b-$quark,
we first check the progenitors of all particles in the jet (except for
the photons). All particles that originate from the decay of one of
the $b-$hadrons $B^0$, $\bar{B}^0$, $B^+$, $B^-$, $B_s^0$,
$\bar{B}_s^0$, $\Lambda_b^0$ or $\Lambda_{\bar{b}}^0$ are marked. If a
jet contains at least one such particle and the pseudorapidity
fulfills $|\eta^{b-jet}| < 2.5$ then the jet is identified as a
$b-$jet at the generator level. Note that the number of $b-$jets could
differ from the number of decaying $b-$hadrons already at this
stage. In particular, the products of multiple $b-$hadrons could end
up in one jet, which is not unlikely if the two corresponding
$b-$quark momenta are nearly parallel to each other. In principle the
decay products of one $b-$hadron could end up in multiple jets, but
this happens only rarely. Finally, a $b-$jet is only tagged with a
$\unit[50]{\%}$ probability. All jets which are not tagged as $b-$jets
are counted as non$-b-$jets. Note that we ignore the possibility that
jets which do not originate from a $b-$quark are tagged as $b-$jets
(false positive tags).

\subsection{Missing Transverse Momentum}

In the simulation the transverse momentum vectors of all measurable
particles are added. The missing transverse momentum
$\vec{\cancel{p}}_T$ is the negative of this sum, $\vec{\cancel{p}}_T
= - \sum_i \vec{\cancel{p}}_{i,T}$. Our event selection includes a cut
on the absolute value of this quantity, $\cancel{p}_T =
|\vec{\cancel{p}}_T|$. 

\subsection{$ H_T $}

$H_T$ is defined as the sum of the transverse momenta of all hard
objects and the absolute value of the missing $p_T$. Hard objects are
all isolated leptons and jets that satisfy the criteria outlined
above.

\section{Selection Cuts}
\label{sec:eventCuts}

As well known, the production of heavy superparticles can be detected
at the LHC on top of SM backgrounds only after cuts have been applied
to reduce these backgrounds. We apply three different sets of
selection cuts, depending on the number and charge of identified
leptons. Note that we include identified $\tau-$jets as leptons for
the purpose of defining these cuts. In the following we outline cuts
for events with at most one lepton, events with exactly two
opposite--charged leptons, and events containing at least two leptons
of the same charge (and possibly some additional lepton(s) of arbitrary charge). All leptons and jets have to satisfy the criteria described
in Appendix \ref{sec:leptonsJetsAndHT}.

\subsection{Events with at Most One Lepton}

We impose the following cuts: 
\begin{itemize}
\item $ \cancel{p}_T > \unit[200]{GeV} $
\item $ H_T >\unit[1000]{GeV} $
\end{itemize}
Additionally there are either exactly two or at least four jets, on
which we apply the following cuts:

\subsubsection{Two Jets}

This event sample will get contributions from squark pair production
if at least one squark decays directly to the lightest neutralino
(possibly plus very soft particles).

\begin{itemize}
\item $ E^j_T > 300, \, \unit[150]{GeV} $ for the hardest and
  second--hardest jet
\item There are no additional jets with transverse energy $E^j_T >
  \unit[30]{GeV}$ and no $b-$jets; this is intended to reduce
  backgrounds from top production
\item $m^{jj} > \unit[200]{GeV} $
\item If there is exactly one lepton then its transverse mass with the
  missing transverse momentum in the event should fulfill $
  m_T(\vec{p}^{\, l}, \, \cancel{\vec{p}}_T) > \unit[80]{GeV}$; this
  suppresses backgrounds where the missing $p_T$ comes from $W^{\pm}
  \rightarrow l^{\pm} \nu_l $ decays. This event sample will get
  contributions from squark pair production where exactly one squark
  decays into a chargino which in turn decays leptonically.
\end{itemize}

\subsubsection{Four or More Jets}

This event sample will get contributions from gluino pair production,
and from squark production in the presence of long decay chains and/or
additional QCD radiation.

\begin{itemize}
\item $ E^j_T > 100, \, 50, \, 50, \, \unit[50]{GeV} $ for the four
  hardest jets
\item Find the smallest invariant mass of two of the four hardest
  jets, $ m^{jj}_{\rm min} $ (there are six combinations). Find the
  smallest invariant mass of three of the four jets, $ m^{3j}_{\rm
    min} $ (there are four combinations). At least one of the
  following three requirements has to be fulfilled, in order to
  suppress $t \bar t$ backgrounds: \subitem Either: $ m^{jj}_{\rm min}
  > \unit[100]{GeV} $ \subitem Or: $ m^{3j}_{\rm min} >
  \unit[200]{GeV} $ \subitem Or: None of the four jets is a $ b-$jet
\item If there is exactly one lepton then its transverse mass with the
  missing transverse momentum in the event should fulfill $
  m_T(\vec{p}^{\, l}, \, \cancel{\vec{p}}_T) > \unit[80]{GeV} $; this
  again suppresses backgrounds where the missing $p_T$ originates from
  a $W$ decay
\end{itemize}

\subsection{Events with Two Opposite--Charged Leptons}

We impose the following cuts:
\begin{itemize}
\item $ \cancel{p}_T > \unit[100]{GeV} $
\item If the two leptons are an $e^+e^-$ or a $\mu^+\mu^-$
  pair their invariant mass should satisfy one of the following
  requirements, which will remove events where the leptons come from a
  $Z \rightarrow l^+ l^-$ decay: 
\subitem Either: $ m^{ll} \leq \unit[75]{GeV} $
\subitem Or: $ m^{ll} \geq \unit[105]{GeV} $
\end{itemize}
In addition, we demand that the event contains either no, at least
three or at least four jets; in the latter two cases we apply
additional cuts in order to suppress the $t \bar t$ background:

\subsubsection{No Jets}

\begin{itemize}
\item There are no jets with $ E^j_T > \unit[30]{GeV} $
\end{itemize}

\subsubsection{Three or More Jets}
\begin{itemize}
\item There are at least three jets with $ E^j_T > 100, \, 100, \,
  \unit[50]{GeV} $ 
\item None of the three highest $ E^j_T $ jets has been tagged as a $b-$jet
\end{itemize}

\subsubsection{Four or More Jets}
\begin{itemize}
\item There are at least four jets, with $ E^j_T > 100, \, 50, \, 50,
  \, \unit[50]{GeV} $ 
\end{itemize}

\subsection{Events with at least Two Leptons of the Same Charge}

We impose the following cuts: 
\begin{itemize}
\item $ \cancel{p}_T > \unit[50]{GeV} $
\item $ \cancel{p}_T > 3 \cdot \sqrt{H_T} $ (in GeV). This is intended
  to suppress events where the missing $p_T$ is due to
  mismeasurements; for example it is not unlikely to find
  $\unit[1]{TeV}$ measured energy accompanied by more than
  $\unit[50]{GeV}$ missing transverse momentum
\item Now consider all lepton pairs with opposite charge but same
  flavor, i.e. $e^+e^-, \, \mu^+\mu^-$ or $\tau^+\tau^-$.
  If there is {\em no} such lepton pair, the event is accepted.
  Otherwise we impose the following two additional cuts:

  \subitem If the event contains exactly three charged leptons, define
  the third lepton $l_3$ as the one {\em not} counted in the
  opposite--charged same--flavored lepton pair. The event then has to
  satisfy $m_T(\vec{p}_{l_3}, \cancel{\vec{p}}_T) > \unit[80]{GeV} $.
  If all three leptons are of the same flavor, there are two possible
  $l^+l^-$ pairs, and hence two possible choices for $l_3$; in this
  case this last cut has to be satisfied for both choices.  
  \subitem In addition, each $l^+l^-$ pair must be an $e^+e^-$ or
  $\mu^+\mu^-$ pair with invariant mass below $\unit[75]{GeV}$ or
  above $\unit[105]{GeV}$ (to suppress $Z \rightarrow l^+l^-$
  backgrounds), or the event must contain at least two jets with $
  E^j_T > 100, \, \unit[100]{GeV} $

\end{itemize}

\end{appendix}

\end{document}